\documentclass[twocolumn,showpacs,preprintnumbers,aps,amsmath]{revtex4}
\usepackage{amsmath,amssymb}
\usepackage{latexsym}
\usepackage[dvips]{graphicx}
\usepackage[dvipdfm ,bookmarksnumbered=true,bookmarkstype=toc,bookmarks=true,linkbordercolor={0 0 1}]{hyperref}

\begin{document}

\preprint{EPHOU 09-004 }

\title{Wrapped brane gas as a candidate for Dark Matter}

\author{Masakazu Sano and Hisao Suzuki}

\affiliation{ Department of Physics,  Hokkaido University, Sapporo, Hokkaido 060-0810 Japan}

\begin{abstract}
We consider  brane gas models based on  type II string theories and analyze the mass, the Ramond-Ramond charge and the charge on moduli fluctuations of branes wrapping over cycles of a compactified space in the four-dimensional Einstein frame. A six-dimensional torus and Calabi-Yau threefolds are considered for the Kaluza-Klein reduction. A large volume of the compactified space and a weak string coupling gives rise to point particles of the wrapped branes which have a light mass and a small charge of the Ramond-Ramond flux and of the moduli fluctuations, while the particles become very heavy in the string frame. 
We find that the masses and the charges satisfy the sea-saw like dual relations which become time-independent in the four-dimensional Einstein frame. 
\end{abstract}


\maketitle

\section{Introduction}

The string gas model \cite{bio_BV}  is one of interesting scenarios in string cosmologies 
(see \cite{bio_BW2, bio_B2, bio_MS, bio_B, bio_BM2} for recent reviews). 
The simplest of such model treats the space as $T^9$. If many strings wrap over cycles of a nine-dimensional torus, $T^{9}$ at a very early universe,  the cosmological singularity can be resolved by T-duality \cite{bio_TV}. 
The string gas models also  provide an intriguing idea on a realization of 
the large three-dimensional space by annihilation of the strings, 
while the scale of the six-dimensional torus is stabilized by the tension of remaining strings. 
The brane gas models \cite{bio_ABE} are  extension of the idea and have been considered a cosmological role of windings of D-branes.

Although these gas models are very interesting, one of the serious problems is the moduli stability.  This problem is a serious obstacle to build 
String/Brane gas models consistent with a realistic phenomenology and cosmology.
This is resulting from massless scalar fields appear on the three dimensions without the moduli stabilization. 
There are many interesting works on the moduli stabilization  
\cite{bio_WB, bio_KR, bio_BBST1, bio_BW, bio_W2, bio_BP1, bio_BP2, bio_P2, 
bio_KS, bio_CWB, bio_C1, bio_BC, bio_BBC, bio_ET, bio_CW, bio_DFB}. 
  Among the stability of many moduli fields, the stabilization of the dilaton field has been difficult in brane gas models.  
Recently, we have shown that  electric fields on D-brane, windings of NS$5$-brane and Kaluza-Klein monopole (KK$5$-monopole) can 
stabilize radial moduli fields of a six-dimensional torus $T^{6}$ and the dilaton field  simultaneously  in
 brane gas models based on type II string theories \cite{bio_SS2}.

According to the result of \cite{bio_SS2}, gases of branes  
wrapping over only cycles of $T^{6}$ make 
the energy density of a pressureless matter, namely the energy density is inversely proportional to the third power of the scale factor. This implies that we observe those wrapped branes as point particles on the three-dimensional space. 
This fact leads to the appearance of the dark matter of the wrapped branes, because wrapped branes are not ordinary matters. 
 The possibility of the dark matter of the wrapped branes has been investigated in various models.  
The appearance of light particles of the wrapped branes in the string frame, 
if the string scale, $\sqrt{\alpha'}$ is larger than the Planck length in Ref.\cite{bio_SW} .
The large scale structure and the accelerating phase were discussed in \cite{bio_GP, bio_GP2, bio_BBMM, bio_BW, bio_NGP}
without an influence of the Ramond-Ramond (RR) flux, although D-branes have the RR charges. 
However, the mass and the charge of RR flux are not  estimated simultaneously in the four-dimensional Einstein frame which is 
required for obtaining realistic models.

We take the four-dimensional Einstein frame, 
when we consider the time-independent four-dimensional gravitational constant after the Kaluza-Klein reduction. 
In this frame, the realization of light wrapped branes is not trivial, 
since it is known that a large volume of the six-dimensional torus and a weak string coupling give rise to 
  heavy wrapped branes in the string frame for a small string length \cite{bio_P}.

The purpose of this paper is to study wrapped branes 
in the brane gas cosmology based on type II string theories as dark matter, using the effective field theory of brane gas models.  
We estimate the mass and the charge on the RR flux and on moduli fluctuations.  These fluctuations also  provide forces between the branes. We find some models where branes can be realized as dust particles whose masses can be around  10TeV scales.
We take the string scale as $1/\sqrt{\alpha'} \sim m_{\text{Planck}}$ and consider a compactification with $T^{6}$ and 
the Calabi-Yau threefold (CY$_{3}$), 
using the four-dimensional Einstein frame which is the key idea to obtain a light mass and a weak interaction. 

The description of the effective field theory requires a large volume of a compactified space and a weak string coupling, 
because quantum corrections should be suppressed and the perturbative expansions should be valid. 
In the string frame, those conditions lead to the appearance of heavy wrapped branes, 
as the world volume action of the brane is proportional to its volume and the inverse of the string coupling.  
 It seems apparent that the dark matter candidate of the wrapped branes may be difficult in this frame.
However, in the four-dimensional Einstein frame, we will show that there are cases where 
the wrapped branes obtain the light mass and the small charge on the RR flux and on the moduli fluctuations
 under the large volume and the weak string coupling.  
We will show that the effective masses and charges satisfy the sea-saw like  dual relations which are time-independent, 
while each quantity depends on the time variable through moduli fields. 
The dual relations are one of reasons for the existence of the various light wrapped branes with the small charges. 

From the phenomenological point of view, the estimation of the mass and charge is mainly done by  
controlling scales of the compactified space by hand.  
However, in this paper, the string scale is of the order of the Planck scale 
and then the existence of light particles of wrapped branes 
is quite non-trivial under the large volume. 
In one model,  we find that the mass of the D$0$-brane is of the order of  $\mathcal{O}(10)$ TeV, 
if we take the six-dimensional compactified volume as $( \mathcal{O}(10^{5}) \times 2 \pi \sqrt{\alpha'})^{6}$ 
where $\sqrt{\alpha'}$ is the string length and holds $m_{\text{Planck}} \sim 1/\sqrt{\alpha'}$ in this paper. 
The square of the effective  charges of the RR flux and of the moduli fluctuations 
is of order  $\mathcal{O}(10^{-30})$ by the same scale of the compactification.  
Then, the D$0$-branes is a possible candidate for the dark matter at a late time.   

We  also consider the masses of wrapped branes by 
a D$1$-KK$5$ brane gas systems \cite{bio_SS2}, because the system can be stabilized by the scale of the $T^{6}$ and the dilaton field.    
It is not trivial that both  the weak string coupling and the large scale 
of the compactification are realized  simultaneously in the specific model after the moduli stabilization. 
We find such realization of those conditions exists
if the number density of D$1$ and KK$5$ satisfy a specific condition which is given later.   
In the case of the Calabi-Yau compactification, we cannot prove 
the moduli stabilization in this model, however the dual relation is also 
satisfied and then we can show that light particles are generated for 
the large volume of the CY$_{3}$.   

This paper is organized as follows. 
In section II, we will give the definition of the four-dimensional Einstein frame. 
In section III, we will consider the effective world volume action  of wrapped branes in the four-dimensional Einstein frame. 
In section IV, we will show the electric magnetic dual relation between masses of the wrapped branes. 
The dual relation explains the existence of the light wrapped branes. 
In section V, we  will see a case in which the electric-magnetic dual relation of  wrapped branes is satisfied in 
the CY$_{3}$ compactification. 
In section VI, we will investigate the effective  RR charge in the four-dimensional Einstein frame. 
The effective  coupling of D$p$- and  D$(6-p)$-brane also satisfies the electric-magnetic dual relation. 
In section VII, we will comment on the charge of fluctuations of the moduli fields, 
because those fluctuations also give interactions 
between various branes. 
In section VIII, we would like to analyze an explicit model constructed by a D$1$-KK$5$ brane gas system. 
Sec. IX will be devoted to the summary and some discussions.

\section{Four-dimensional Einstein frame}

In this section, we shall define the four-dimensional Einstein frame which 
gives the time-independent four-dimensional gravitational constant after a compactification. 
In general, the Kaluza-Klein reduction gives rise to a coupling between the four-dimensional 
Einstein-Hilbert term and various moduli fields. 
If those moduli fields are functions of coordinates of the compactified space, 
we can integral out moduli fields and renormalize the factor into the higher dimensional gravitational constant. 
However, in cosmologies, moduli fields depend on the time and, therefore,  we cannot integral out the moduli fields completely.   
The time-independent Newton constant  can be easily realized in the four-dimensional Einstein frame which is useful to analyze a behavior of scale factors and various fields.

We consider a homogeneous ten-dimensional metric and a six-dimensional torus, $T^{6}$ as a compactified space.
 The scale factors are functions of the time coordinate. 
The $T^{6}$ has six scale factors corresponding to six cycles. 
The string-frame metric is given by the following equation: 
\begin{equation}
ds_{10}^{2}=-e^{2 \lambda_{0}(t)} dt^{2}+e^{2 \lambda(t)} d{\bf x}^2+\sum_{m=4}^{9} e^{2 \lambda_{m}(t)}(dy^{m})^2 \label{e1} 
\end{equation}
where $d\mathbf{x}^{2}\equiv \sum_{i=1}^{3}(dx^{i})^{2}$ is the line element of the flat three-dimensional space, 
$R^{3}$ or $T^{3}$ and the cycle of $T^{6}$ is defined as
\begin{equation}
0\leq y^{m} \leq 2 \pi \sqrt{\alpha'}.  \label{e2}
\end{equation}
$\alpha'$ is related with a string length as $l_{\text{s}} \sim \sqrt{\alpha'}$. 
The scale factor $e^{\lambda_{m}(t)}$ describes the scaling of the cycle defined by (\ref{e2}). 
The four-dimensional Einstein frame is defined by the following transformation:
\begin{equation}
\begin{split}
\lambda _{0}(t)&=n(t)+\beta(t) ,\qquad  \beta(t)\equiv \phi(t)-\frac{1}{2}\overline{\lambda }(t), \\
\lambda (t)&=A(t)+\beta(t) ,\qquad \overline{\lambda }(t)\equiv \sum_{m=4}^{9}\lambda _{m}(t)  
\end{split} \label{e3}
\end{equation}
where $\phi(t)$ is dilaton field. 
$e^{n(t)}$ and $e^{A(t)}$ are the lapse function and the scale factor of the three-dimensional space, respectively. 
For the Einstein frame the proper time of the four-dimensional space-time is defined by $e^{n(t)}=1$ which 
is not equivalent to the proper time of the ten-dimensional space-time in the string frame because of $e^{\lambda_{0}(t)}=e^{\beta(t)}$.

By (\ref{e3}) 
the dilaton gravity sector of the string frame action becomes  \cite{bio_TV}:
\begin{align}
S_{0}&=\frac{1}{16 \pi G_{10}} \int_{M^{10}} d^{10}X \sqrt{-G} e^{-2\phi(t)} (R+4(\nabla \phi)^{2}) \notag \\
&=\frac{1}{16 \pi G_{4}} \int_{M^{4}} d^{4}x e^{n(t)+3A(t)} \notag \\
& \qquad \times e^{-2n(t)} \Bigl( -6\dot{A}^{2}(t)+2\dot{\beta}^{2}(t)+\sum_{m=4}^{9}\dot{\lambda}_{m}^{2}(t) \Bigr) \label{e4}
\end{align}
where the four-dimensional Newton constant and the ten-dimensional gravitational constant are given by \cite{bio_P} 
\begin{align}
G_{4}&=\frac{G_{10}}{( 2 \pi \sqrt{\alpha'})^{6}}=\frac{\alpha' }{8},  \label{e5} \\
\kappa_{10}^{2}&=8\pi G_{10}=\frac{1}{2}(2 \pi)^{7} \alpha'^{4}.  \label{e_newton}
\end{align}

Eq. (\ref{e4}) shows that the four-dimensional Einstein-Hilbert term which is given by the first term of the second line 
does not involve moduli fields and the dilaton. 
By (\ref{e5}) we find that the string length is of the order of the Planck length.  
An advantage of the new variable $\beta(t)$ is that
the kinetic term of the above action becomes diagonal.

\section{point particles from Wrapped Branes}

In the previous section, we have defined the four-dimensional Einstein frame which gives 
the four-dimensional Einstein-Hilbert term with the time-independent Newton constant and diagonal kinetic terms of moduli fields. We found that the dynamics of the moduli fields is governed by potential terms derived by flux terms and brane sources after the Kaluza-Klein reduction. 
In this section, we will consider the effective action of the wrapped D$p$-brane, NS$5$-brane and KK$5$-monopole, 
using the four-dimensional Einstein frame and derive the energy density of those ingredients which contribute as potential terms for the moduli fields.

We consider branes wrapping over only cycles of $T^{6}$. 
After the compactification, we see those objects as point particles on the three-dimensional space. 
This expectation is motivated by ref. \cite{bio_SS2} which shows 
that wrapped brane gases give the energy density of a pressureless matter. 
In \cite{bio_SS2}, we gave the effective action without a dependence of the velocity along the three-dimensional space. 
We would like to derive the effective action with the velocity on the three-dimensional space. 
The effective action will be used to read off the mass and the charge of 
a wrapped brane in later sections.  
To derive the effective action we assume a cancellation of  total RR charges of the D-branes, because of 
homogeneous distributions of  many D-branes on the nine-dimensional space by the brane gas approximation.   
Thus, we will assume wrapped branes as gases of free particles.

First, we will consider the world volume action of D$p$-brane wrapping over a $p$-dimensional cycle of $T^{6}$. 
If the D$p$-brane wraps over a ($m_{1}, \dots ,m_{p}$)-cycle ($0\leq  p\leq 6$),   
 gauge fields on the D$p$-brane exist along the ($m_{1}, \dots ,m_{p}$)-directions. 
The D$p$-brane can move along transverse directions.  
We assume that the distributions of the branes are homogeneous and that the gauge field is Abelian. Then the gauge potential and transverse coordinates depend only on the time variable, i.e. 
$A_{m_{a}}(t)$, $X^{i}(t)$ ($i=1,\,2,\,3$) and $X^{m_{a}}(t)$ .
We will adopt the coordinate system as $\xi^{0}=t$, $\xi^{m_{a}}=y^{m_{a}}$ ($a=1,\,2,\,\dots ,\,p$).
Then, the Dirac-Born-Infeld  action of D$p$-brane wrapping over the ($m_{1}, \dots ,m_{p}$)-cycle, $\varSigma_{p}$ is given by
\begin{align}
&S_{\text{D}p}^{(m_{1}\cdots m_{p})} \notag \\
=&-T_{p}\int_{R\times \varSigma_{p}}d^{p+1}\xi e^{-\phi(t)}\sqrt{-\det(\gamma_{ab}+2\pi \alpha' F_{ab})} \notag \\
=&-(2\pi \sqrt{ \alpha'} \,)^{p}T_{p}\int dt e^{-\phi(t)+\lambda _{0}(t)+ \sum_{a=1}^{p}\lambda _{m_{a}}(t)}  \notag \\
&\times \Bigl\{ 1 -e^{-2\lambda_{0}(t)+2\lambda(t)} \sum_{i=1}^{3} (\dot{X}^{i}(t) )^{2} \notag \\
&-(2\pi \alpha')^{2}\sum_{a=1}^{p}e^{-2 \lambda _{0}(t)-2\lambda _{m_{a}}(t)}(\dot{A}_{m_{a}}(t))^{2} \notag \\
&-\sum_{b=p+1}^{6}e^{-2 \lambda _{0}(t)+2 \lambda _{m_{b}}(t)} (\dot{X}^{m_{b}}(t))^{2} \Bigr\}^{\frac{1}{2}} \label{e6}
\end{align}
where the tension of the D$p$-brane is given by \cite{bio_P}
\begin{equation}
\begin{split}
T_{p}=(2\pi)^{-p}(\alpha')^{-(p+1)/2}. \label{e7}
\end{split}
\end{equation}
$\sum_{i=1}^{3} (\dot{X}^{i}(t) )^{2}$ denotes the velocity along the three-dimensional space. 
The world volume action is proportional to $\exp(-\phi) \sqrt{\det \gamma_{ab}}$ 
and then the mass becomes heavy for $e^{\phi}\ll 1$ and $e^{\lambda_{m}}\gg 1$.

The above action is defined in the string frame. We have to take the four-dimensional frame 
to extract 
the four-dimensional dynamics. Substituting (\ref{e3}) for (\ref{e6}), we obtain 
\begin{align}
&S_{\text{D}p}^{(m_{1}\cdots m_{p})} \notag \\
=&-(2\pi\sqrt{\alpha' } )^{p}T_{p}\int dt e^{n(t)-\frac{1}{2}\overline{\lambda }(t)+ \sum_{a=1}^{p}\lambda _{m_{a}}(t)}  \notag \\
&\times \Bigl\{ 1 -e^{-2n(t)+2A(t)} \sum_{i=1}^{3} (\dot{X}^{i}(t) )^{2}
-\mathcal{A}(t) \Bigr\}^{\frac{1}{2}} \label{e8}
\end{align}  
where $\mathcal{A}(t)$ is defined by the following equation: 
\begin{align}
\mathcal{A}(t)&\equiv (2\pi \alpha')^{2}\sum_{a=1}^{p}e^{-2\beta(t) -2 n(t)-2\lambda _{m_{a}}(t)}(\dot{A}_{m_{a}}(t))^{2} \notag \\
&+\sum_{b=p+1}^{6}e^{-2 \beta(t)-2 n(t) +2\lambda _{m_{b}}(t)} (\dot{X}^{m_{b}}(t))^{2}. \label{e9}
\end{align}
Compared with (\ref{e8}) to (\ref{e6}), 
we find the basic difference between the string frame and Einstein frame.  
In the string frame, the winding mode of the world volume depends on the dilaton, 
whereas 
the winding mode of D$p$-brane does not have the coupling of the dilaton field in the four-dimensional Einstein frame. 
This fact implies that the winding mode cannot stabilize the dilaton field in the four-dimensional Einstein frame. 
Namely, if we take $\mathcal{A}(t)$ defined by (\ref{e9}), we also cannot stabilize the dilaton, 
since $\mathcal{A}(t)$ includes the dilaton with $e^{-2\beta(t)}$ only. 
Therefore, the D$p$-brane wrapping over cycles of $T^{6}$ 
 cannot stabilize the dilaton field in the four-dimensional Einstein frame, 
using the cosmological background defined by (\ref{e1}).

The wrapped branes can be regarded as the ideal gas based on the dilute gas approximation.  
We can solve the equations of motion of $A_{m_{a}}(t)$ and $X^{m_{b}}(t)$ derived from (\ref{e8}) as follows:  
\begin{align}
&e^{n(t)-\frac{1}{2}\overline{\lambda }(t)+ \sum_{a'=1}^{p}\lambda _{m_{a'}}(t)} \notag \\
&\qquad \times (2\pi \alpha') 
e^{-2\beta(t) -2 n(t)-2\lambda _{m_{a}}(t)} \dot{A}_{m_{a}}(t) \notag \\
=&\frac{f_{m_{a}}}{2\pi \alpha'} 
\Bigl\{ 1 -e^{-2n(t)+2A(t)} \sum_{i=1}^{3} (\dot{X}^{i}(t) )^{2}
-\mathcal{A}(t) \Bigr\}^{\frac{1}{2}}, \label{e10} \\[15pt]
&e^{n(t)-\frac{1}{2}\overline{\lambda  }(t)+ \sum_{a'=1}^{p}\lambda _{m_{a'}}(t)} \notag \\
&\qquad \times e^{-2\beta(t)-2 n(t) +2\lambda _{m_{b}}(t)}\dot{X}^{m_{b}}(t) \notag \\
=&v^{m_{b}} 
\Bigl\{ 1 -e^{-2n(t)+2A(t)} \sum_{i=1}^{3} (\dot{X}^{i}(t) )^{2}
-\mathcal{A}(t) \Bigr\}^{\frac{1}{2}} \label{e11}
\end{align}
where we have the equations of motions and $f_{m_{a}} $ and 
$v^{m_{b}}$ are constants of integration. 
$f_{m_{a}}$ and $v^{m_{b}}$ are integers and not continuouse numbers, because of a cyclicity of the $T^{6}$. 
For exmaple, $v^{m_{b}}$ is related a conjugate momentum $P_{\text{D}p,\,m_{b}}^{m_{1}\cdots m_{p}} \equiv \partial \mathcal{L}_{\text{D}p}^{m_{1} \cdots m_{m_{p}}}/\partial \dot{X}^{m_{b}}$ as 
$P_{\text{D}p,\,m_{b}}^{m_{1}\cdots m_{p}}=v^{m_{b}}/\sqrt{\alpha'}$ by (\ref{e11}). 
The conjugate momentum must be an integer on a cycle of the $T^{6}$ therefore $v^{m_{b}}$ is an integer. 
By the T-dulaity $f_{m_{a}}$ and $v^{m_{b}}$ is mapped each 
other \cite{bio_SS2} and then $f_{m_{a}}$ must be also an integer. 
The similar property on a quantized momentum is satisfied for 
a fundamental string as shown in Appendix \ref{appendixC}.
 
We can solve the equations (\ref{e10}) and (\ref{e11}) on $\mathcal{A}(t)$: 
\begin{equation}
\begin{split}
\mathcal{A}(t)&=\frac{  e^{\overline{\lambda}-2\sum_{a'=1}^{p}\lambda _{m_{a'}}(t)} \widetilde{\mathcal{A}}(t) }
{1+ e^{\overline{\lambda}-2\sum_{a'=1}^{p}\lambda _{m_{a'}}(t)} \widetilde{\mathcal{A}}(t)  }   \\
& \times \Bigl\{ 1 -e^{-2n(t)+2A(t)} \sum_{i=1}^{3} (\dot{X}^{i}(t) )^{2} \Bigr\}, \\
 \widetilde{\mathcal{A}}(t)&\equiv \sum_{a=1}^{p}e^{2\beta (t)+2 \lambda _{m_{a}}(t)} 
\Bigg( \frac{  f_{m_{a}} }{2\pi \alpha'} \Biggr)^{2}  \\
&+\sum_{b=p+1}^{6}e^{2\beta(t)-2 \lambda _{m_{b}}(t)} (v^{m_{b}} )^{2} .
\label{e12}
\end{split}
\end{equation}

In the four-dimensional Einstein frame, the potential term is derived by 
\begin{align}
u_{\text{D}p}^{(m_{1}\cdots m_{p})} \equiv-T^{0}_{~0}&=\frac{2 g^{00}}{\sqrt{-g_{4}(x)}}\frac{\delta \mathcal{L}_{\text{D}p}^{(m_{1}\cdots m_{p})}}{\delta g^{00}} \notag \\
&=-e^{-n(t)-3A(t)}\frac{\delta \mathcal{L}_{\text{D}p}^{(m_{1}\cdots m_{p})}}{\delta n(t)} . \label{e13}
\end{align}
Using (\ref{e8}), (\ref{e12}) and (\ref{e13}), 
we obtain the potential of the D$p$-brane wrapping over the ($m_{1} \cdots m_{p}$)-cycle:
\begin{align}
u_{\text{D}p}^{(m_{1}\cdots m_{p})} 
=& e^{-3A(t)} (2\pi \sqrt{\alpha' } )^{p}T_{p}  \notag \\
&\times \Bigl\{ e^{ -\overline{\lambda }(t)+ 2 \sum_{a=1}^{p}\lambda _{m_{a}}(t)} +  
 \widetilde{\mathcal{A}}(t)  \Bigr\}^{\frac{1}{2}} \notag \\
&\times  \Bigl\{ 1 -e^{-2n(t)+2A(t)} \sum_{i=1}^{3} (\dot{X}^{i}(t) )^{2} \Bigr\}^{-\frac{1}{2}}. \label{e14}
\end{align}
This corresponds to the energy density of a relativistic particle  with the mass given by 
\begin{equation}
\begin{split}
m^{(m_{1}\cdots m_{p})}_{p}=&(2\pi \sqrt{\alpha' } )^{p}T_{p} \\
&\times  \Bigl\{ e^{ -\overline{\lambda }(t)+ 2 \sum_{a=1}^{p}\lambda _{m_{a}}(t)} +  
 \widetilde{\mathcal{A}}(t)  \Bigr\}^{\frac{1}{2}}.  
\end{split}\label{e15}
\end{equation}
\begin{figure}[b]
\begin{center}
\includegraphics[width=6cm,clip]{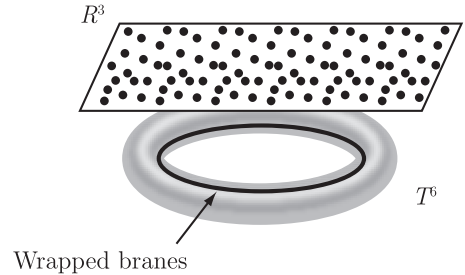}
\end{center}
\caption{Wrapped brane gases can be seen as point particles on three dimensions. 
Those point particles decay as the pressureless dust with $e^{-3A(t)}$.}
\label{fig1}
\end{figure}
Thus, we see  D$p$-branes  behave as  point particles 
on the three-dimensional space as Fig.\ref{fig1}.  The world volume 
decays with $e^{-3A(t)}$ as the pressureless matter.

\begin{figure}[t]
\begin{center}
\includegraphics[width=6cm,clip]{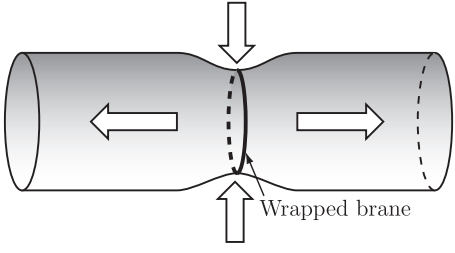}
\end{center}
\caption{A cycle where branes wrap is bound by the tension of branes. 
Cycles where branes do not wrap are stretched like a rubber tube in the four-dimensional Einstein frame. 
The cycles are homogeneously bound under the brane gas approximation, 
although this figure shows one of branes as an example.}
\label{fig2}
\end{figure}

The mass given by (\ref{e15}) has the following role as a potential of the moduli fields. 
The winding part of (\ref{e15}) is proportional to
\begin{equation} 
\exp \frac{1}{2}\Biggl( +\sum_{a=1}^{p}\lambda_{m_{a}}(t)-\sum_{b=p+1}^{6}\lambda_{m_{b}}(t)  \Biggr). \notag 
\end{equation} 
This indicates that the wrapped D$p$-brane binds the ($m_{1} \cdots m_{p}$)-cycle where the D$p$-brane expands and  
the ($m_{p+1}\cdots m_{6}$)-direction is stretched like a rubber band as in Fig.\ref{fig2}.
Similarly, by (\ref{e12}) it is found that the electric fields in $\widetilde{\mathcal{A}}(t)$ also bind the ($m_{1} \cdots m_{p}$)-cycle. 
 The transverse velocities in $\widetilde{\mathcal{A}}(t)$ make a pressure bringing the expansion to the transverse directions.

We will derive  the action of the point particle for NS$5$-brane and KK$5$-monopole in type II string theory which has five spatial dimensions.
In this paper, we consider only winding modes of NS$5$-brane and KK$5$-monopole moving along the three-dimensional space. 
The world volume action of KK$5$-monopole wrapping over the ($m_{1}\cdots m_{5}$)-cycle is given by \cite{bio_BJO, bio_EJL} 
\begin{align}
&S^{(m_{1}\cdots m_{5})}_{\text{KK}5} 
=-T_{\text{KK}5}\int_{R\times \varSigma_{5}} d^{6}\xi e^{-2\phi(t)} k^{2} \sqrt{-\det \widetilde{\gamma}_{ab}} \notag \\
=&-(2\pi \sqrt{\alpha' } )^{5}T_{\text{KK}5} \int dt e^{-2\phi(t)+2\lambda _{6}(t)+\lambda _{0}(t)+\sum_{a=1}^{5}\lambda _{m_{a}}(t)} \notag \\
& \times \Bigl\{ 1 -e^{-2\lambda_{0}(t)+2\lambda(t)} \sum_{i=1}^{3} (\dot{X}^{i}(t) )^{2} \Bigr\}^{\frac{1}{2}}, \notag \\
&\widetilde{\gamma}_{ab} \equiv \frac{\partial X^{m}}{\partial \xi^{a}} \frac{\partial X^{m}}{\partial \xi^{a}} 
( G_{mn}-k^{-2}k_{m}k_{n}  ).  \label{e16} 
\end{align}
where $k^{m}\equiv \delta^{m m_{6}}$ is the killing vector of the $S^{1}$ isometry along the ($m_{6}$)-cycle and 
$k^{2}\equiv G_{mn}k^{m}k^{n}=\exp{(2 \lambda_{m_{6}}(t))}$. 
We also consider the  NS$5$-brane. 
The world volume action of the NS$5$-brane \cite{bio_BLO1} wrapping over ($m_{1}\cdots m_{5}$)-cycles is given by 
\begin{align}
&S^{(m_{1}\cdots m_{5})}_{\text{NS}5}
=-T_{\text{NS}5}\int_{R\times \varSigma_{5}} d^{6}\xi e^{-2\phi(t)} \sqrt{-\det\gamma_{ab}} \notag \\
=&-(2\pi \sqrt{\alpha' })^{5}T_{\text{NS}5} \int dt e^{-2\phi(t)+\lambda _{0}(t)+\sum_{a=1}^{5}\lambda _{m_{a}}(t)} \notag \\
& \times \Bigl\{ 1 -e^{-2\lambda_{0}(t)+2\lambda(t)} \sum_{i=1}^{3} (\dot{X}^{i}(t) )^{2} \Bigr\}^{\frac{1}{2}}. \label{e17}
\end{align}

 We consider the transformation of fields as in  (\ref{e3}) 
to obtain the potential term of the NS$5$-brane and the KK$5$-monopole in the four-dimensional Einstein frame.  
Substituting (\ref{e3}) for (\ref{e16}) and (\ref{e17}) and replacing $\mathcal{L}_{\text{D}p}^{(m_{1}\cdots m_{p})}$ with $\mathcal{L}_{\text{NS}5/\text{KK}5}^{(m_{1}\cdots m_{5})}$ in 
(\ref{e13}), the energy density of the NS$5$-brane and the KK$5$-monopole are given by 
\begin{align}
u_{\text{NS}5}^{(m_{1}\cdots m_{5})}&=e^{-3A(t)}(2\pi\sqrt{\alpha' } )^{5}T_{\text{NS}5} \notag \\
&\qquad  \times e^{-\beta(t)-\overline{\lambda  }(t)+\sum_{a=1}^{5}\lambda _{m_{a}}(t)} \notag \\
&\qquad \times \Bigl\{ 1 -e^{-2n(t)+2A(t)} \sum_{i=1}^{3} (\dot{X}^{i}(t) )^{2} \Bigr\}^{-\frac{1}{2}}, \label{e18} \\
u_{\text{KK}5}^{(m_{1}\cdots m_{5})}&=e^{-3A(t)}(2\pi\sqrt{\alpha' })^{5}T_{\text{KK}5} \notag \\
&\qquad  \times e^{-\beta(t)-\overline{\lambda  }(t)+2\lambda _{m_{6}}+\sum_{a=1}^{5}\lambda _{m_{a}}(t)} \notag \\
&\qquad \times \Bigl\{ 1 -e^{-2n(t)+2A(t)} \sum_{i=1}^{3} (\dot{X}^{i}(t) )^{2} \Bigr\}^{-\frac{1}{2}}, \label{e19}
\end{align}
where  $T_{\text{D}5}=T_{\text{NS}5}=T_{\text{KK}5}$ \cite{bio_EJL}. 
The mass of the particles is given by
\begin{align}
m_{\text{NS}5}^{(m_{1}\cdots m_{5})}&=(2\pi\sqrt{\alpha' } )^{5}T_{\text{NS}5}  \times e^{-\beta(t)-\lambda _{m_{6}}(t)}, \label{e20} \\
m_{\text{KK}5}^{(m_{1}\cdots m_{5})}&=(2\pi\sqrt{\alpha' })^{5}T_{\text{KK}5}  \times e^{-\beta(t)+\lambda _{m_{6}}(t)}. \label{e21}
\end{align}
It is expected that the world volume of NS$5$ and KK$5$ contributes to the stabilization of the dilaton, 
since those objects have the opposite dependence on the dilaton, compared with (\ref{e15}).  
In fact, \cite{bio_SS2}  have suggested that a D$1$-KK$5$ brane gas model stabilizes the dilaton field as well as the radial moduli fields, 
simultaneously.

\section{Mass of wrapped branes in the four-dimensional Einstein frame}

In section III, the effective mass of the wrapped brane is derived in the four-dimensional Einstein frame. 
In this section, we would like to consider a behavior of the mass derived in the previous section, taking a large scale of the $T^{6}$ and 
a weak string coupling related with the dilaton field as $g_{s}=\exp(\phi)$.  
The condition is required to consider an effective field theory. 
It is not necessarily correct that D$p$-brane is very heavy with a large volume and a weak string coupling,  
since the world volume action has the nontrivial coupling to the dilaton and the scale of $T^{6}$ in the four-dimensional Einstein frame. 
We take the string scale as $1/\sqrt{\alpha'} \sim m_{\text{Planck}} \sim \mathcal{O}(10^{19})$ GeV to analyze the mass. 

First, we will consider  winding modes of branes and  
$e^{\lambda_{m}(t)}=e^{\lambda'(t)}$. 
By  (\ref{e15}), (\ref{e20}) and (\ref{e21}) the mass of branes is as follows: 
\begin{equation}
\begin{split}
&m^{(m_{1}\cdots m_{p})}_{p,\,\text{winding}}= 
 \frac{1}{\sqrt{\alpha'}} \times e^{ (p-3)\lambda'(t)}, \\
&m_{\text{NS}5,\,\text{winding}}^{(m_{1}\cdots m_{5})} =\frac{1}{\sqrt{\alpha'}} \times e^{ - \phi (t) +2\lambda'(t)}, \\
&m_{\text{KK}5,\,\text{winding}}^{(m_{1}\cdots m_{5})} = \frac{1}{\sqrt{\alpha'}} \times e^{ - \phi (t) +4\lambda'(t)}. \label{en1}
\end{split}
\end{equation}
We find that the mass of NS$5$ and KK$5$ becomes very heavy for $e^{\lambda'} \gg 1$ and 
$e^{\phi}\ll 1$. In the case of $p\geq 4$, the mass of D$p$-brane also becomes heavy. 
The D$3$-brane has the Planck mass. 
On the other hand, for $p\leq 2$, the D$p$-brane has the light mass 
under the condition of $e^{\lambda'} \gg 1$ and $e^{\phi}\ll 1$. 
For example, we will consider $\exp(\lambda') \sim \mathcal{O}(10^{5})$ which implies that 
the compactification is around GUT scale, 
\begin{equation} 
V_{T^{6}}=( 2 \pi \sqrt{\alpha'} \exp (\lambda') )^{6} \sim ( \mathcal{O}(10^{-14}) \text{GeV}^{-1})^{6}. \notag 
\end{equation} 
The mass of the D$p$-brane is of the order of 
\begin{equation}
m^{(m_{1}\cdots m_{p})}_{p,\,\text{winding}} \sim \mathcal{O}(10^{5(p-3)+19}) \text{ GeV}. \label{mass_torus_scaling}
\end{equation} 
The mass ranges from $\mathcal{O}(10)$ TeV for $p=0$ to $\mathcal{O}(10^{14})$ GeV for $p=2$. 
This is the non-trivial result, 
since the string frame gives rise to heavy branes for the large volume. 
We should include those light states in string cosmologies. 
Taking the four-dimensional Einstein frame gives the interesting result for the mass of wrapped D-branes.

Secondly, we will consider the anisotropic case, $\lambda_{m}\neq \lambda_{n}$ satisfying $\exp ( \lambda_{m}) \gg1$ 
for the large volume condition. 
By (\ref{e15}) the masses of the wrapped D$p$- and D$(6-p)$-brane can be represented as 
\begin{equation}
\begin{split}
m^{(m_{1}\cdots m_{p})}_{p,\,\text{winding}}&= 
 \frac{1}{\sqrt{\alpha'}}  \exp 
\Biggl( \sum_{a=1}^{p}\frac{\lambda_{m_{a}}(t)}{2}
-\sum_{b=p+1}^{6}\frac{\lambda_{m_{b}}(t)}{2} \Biggr), \\
m^{(m_{p+1}\cdots m_{6})}_{6-p,\,\text{winding}}&=\frac{1}{\sqrt{\alpha'}}  \exp 
\Biggl( \sum_{a=p+1}^{6}\frac{\lambda_{m_{a}}(t)}{2}-\sum_{b=1}^{p}\frac{\lambda_{m_{b}}(t)}{2} \Biggr).
\end{split}  \label{dual0}
\end{equation}
Those masses explicitly satisfy the following electric-magnetic dual relation:
\begin{equation}
m^{(m_{1}\cdots m_{p})}_{p,\,\text{winding}}  \times m^{(m_{p+1}\cdots m_{6})}_{6-p,\,\text{winding}}=
\Bigl(\frac{1}{\sqrt{\alpha'}} \Bigr)^{2}. \label{dual1}
\end{equation}
The right hand side of the above equation is independent on scale factors, 
while  masses depend on the time. 
Thus,  masses of D$p$- and D$(6-p)$-brane are not independent each other and satisfy a sea-saw like condition.
 A D$p$-brane becomes light when a D$(6-p)$-brane has a heavy mass and vice versa. 
The electric-magnetic dual relation is one of the reason for this scaling property. 
The behavior of the mass can be classified into three parts by a dynamics of the moduli fields as follows. 
\begin{enumerate}
\item[(1)] $\sum_{a=1}^{p}\lambda_{m_{a}}(t)-\sum_{b=p+1}^{6}\lambda_{m_{b}}(t)>0$: 

the mass of D$p$-brane is heavy and D$(6-p)$ becomes light. 
\item[(2)] $\sum_{a=1}^{p}\lambda_{m_{a}}(t)-\sum_{b=p+1}^{6}\lambda_{m_{b}}(t)=0$: 

the masses of D$p$- and D$(6-p)$-brane are of order $\mathcal{O}(1/\sqrt{\alpha'})\sim \mathcal{O}(m_{\text{Planck}})$. 
\item[(3)] $\sum_{a=1}^{p}\lambda_{m_{a}}(t)-\sum_{b=p+1}^{6}\lambda_{m_{b}}(t)<0$:

the mass of D$(6-p)$-brane is heavy and D$p$-brane becomes light. 
\end{enumerate}

A comment is  in order about the dimensionality of space-times. 
The time-independent electric-magnetic dual relation (\ref{dual1}) is realized only in the four-dimensional Einstein frame. 
For example, the mass of wrapped D$p$-brane in the string frame is given by 
\begin{equation}
m^{(\text{string}|m_{1}\cdots m_{p})}_{p,\,\text{winding}}=\frac{1}{\sqrt{\alpha'}}
 e^{-\phi (t)+\lambda_{m_{1}}(t)+\cdots +\lambda_{m_{p}}(t) } \notag 
\end{equation}
 and the dual relation is \cite{bio_OP}
\begin{align}
m^{(\text{string}|m_{1}\cdots m_{p})}_{p,\,\text{winding}} \times &  m^{(\text{string}|m_{p+1}\cdots m_{6})}_{6-p,\,\text{winding}} \notag \\
&=\Bigl(\frac{1}{\sqrt{\alpha'}} \Bigr)^{2}
\exp \Bigl( -2\phi (t)+\overline{\lambda} (t) \Bigr). \label{dual2}
\end{align}
This relation explicitly depends on time.
If the right hand side of (\ref{dual2}) becomes large, 
each mass of wrapped branes can take a large value. 
Similarly, using the ($d+1$)-dimensional Einstein frame given by (\ref{app2}), 
the mass of the wrapped D$p$-brane is given by 
\begin{align}
&m^{(d+1|m_{1}\cdots m_{p})}_{p,\,\text{winding}}=\frac{1}{\sqrt{\alpha'}}  \notag \\
&\times \exp \Biggl(  \frac{3-d}{d-1}\phi (t)+\sum_{a=1}^{p}\lambda_{m_{a}}(t)-\frac{1}{d-1}\sum_{m=d+1}^{9}\lambda_{m}(t) \Biggr) \label{dual3}
\end{align}
and the dual relation is 
\begin{align}
&m^{(d+1|m_{1}\cdots m_{p})}_{p,\,\text{winding}} \times  m^{(d+1|m_{p+1}\cdots m_{10-(d+1)})}_{10-(d+1)-p,\,\text{winding}}  \notag \\
&\quad =\Bigl(\frac{1}{\sqrt{\alpha'}} \Bigr)^{2}
\exp \frac{d-3}{d-1}\Biggl(   -2\phi (t) +\sum_{m=d+1}^{9}\lambda_{m} (t)  \Biggr). \label{dual4}
\end{align}
The right hand side also involves time-dependent scale factors.  
Therefore, the four-dimensional Einstein frame is a special case where 
the electric-magnetic dual relation becomes scale-free.  

By  the S- and T-duality, it is expected that the mass of the wrapped NS$5$-brane and KK$5$-monopole has a dual relation as (\ref{dual1}).  
Using S-dual rule in the four-dimensional Einstein frame (\ref{app5}) and (\ref{app6}), 
the winding modes of D$1$- and D$5$-brane are mapped to the winding modes of fundamental string given by the first term of 
(\ref{app11}) with $w^{m_{a}}=1$ and NS$5$-brane, respectively. 
Then wrapped fundamental string and NS$5$-brane satisfy the following dual relation: 
\begin{equation}
m^{(m_{1})}_{\text{F}1,\,\text{winding}} \times m^{(m_{2}\cdots m_{6})}_{\text{NS}5}=
\Bigl(\frac{1}{\sqrt{\alpha'}} \Bigr)^{2} \label{dual_5}
\end{equation}
where we have used (\ref{e20}) and (\ref{app11}). According to T-dual along ($m_{1}$)-cycle 
($\lambda_{m_{1}} \rightarrow  -\lambda_{m_{1}}$, $\phi \rightarrow \phi-\lambda_{m_{1}}$), 
the mass of wrapped NS$5$-brane is mapped to 
the mass of KK$5$-monopole as $m^{(m_{2}\cdots m_{6})}_{\text{NS}5} \rightarrow m^{(m_{2}\cdots m_{6})}_{\text{KK}5}$ \cite{bio_SS2}.
According to T-dual along ($m_{1}$)-cycle 
($\lambda_{m_{1}} \rightarrow  -\lambda_{m_{1}}$, $\phi \rightarrow \phi-\lambda_{m_{1}}$), 
the mass of wrapped NS$5$-brane is mapped to 
the mass of KK$5$-monopole as $m^{(m_{2}\cdots m_{6})}_{\text{NS}5} \rightarrow m^{(m_{2}\cdots m_{6})}_{\text{KK}5}$ \cite{bio_SS2}.
  The winding mode of the fundamental string is mapped to the momentum mode as 
$m^{(m_{1})}_{\text{F}1,\,\text{winding}} \rightarrow m^{(m_{1})}_{\text{F}1,\,\text{momentum}}=\frac{1}{\sqrt{\alpha'}} 
 e^{\beta (t)-\lambda_{m_{1}}(t)}$, ($n_{m_{1}}=1$).  Therefore we obtain 
\begin{equation}
m^{(m_{1})}_{\text{F}1,\,\text{momentum}} \times m^{(m_{2}\cdots m_{6})}_{\text{KK}5}=
\Bigl(\frac{1}{\sqrt{\alpha'}} \Bigr)^{2}. \label{dual_6}
\end{equation}
The right hand side of (\ref{dual_5}) and (\ref{dual_6}) is scale free in the four-dimensional Einstein frame.  
For example, if $m^{(m_{1})}_{\text{F}1,\,\text{winding}}$ and $m^{(m_{1})}_{\text{F}1,\,\text{momentum}}$ are lighter than the Planck mass, 
the mass of the the wrapped NS$5$-brane and the wrapped KK$5$-monopole becomes very heavy. 

The dual relations (\ref{dual_5}) and (\ref{dual_6})
are related with the stabilization of the dilaton field. 
The winding and the momentum mode of the fundamental string is proportional to 
$e^{+\beta}$. 
By dual relations, the mass of NS$5$ and KK$5$ must have the dependence of $e^{-\beta}$ on the dilaton field.  
NS$5$ and KK$5$ are essential for the stabilization of the dilaton \cite{bio_SS2}. 
Thus the dual relation of wrapped branes plays an important role
for the moduli stabilization.

\section{mass of wrapped branes with Calabi-Yau threefold}

In the previous section, we have found that there are cases where the mass of the brane wrapping over cycles of the $T^{6}$
 becomes light in the four-dimensional Einstein frame. 
One may expect a light wrapped branes for the CY$_{3}$ compactification.  
In this section, we will investigate masses of D$p$-branes wrapping over cycles of the CY$_{3}$, taking the four-dimensional Einstein frame. 
To analyze the mass we control scales of the Calabi-Yau by hand 
because we cannot show the moduli stabilization of the Calabi-Yau in this paper. 
It is also interesting to know whether  masses satisfy the electric-magnetic dual relation found in the previous section.

The CY$_{3}$ has real six dimensions, 2-, 3- and 4-cycles. 
D$0$-, D$2$-, D$3$-, D$4$- and  D$6$-brane can live in the CY$_{3}$, 
while D$1$-, D$5$-, NS$5$-brane and KK$5$-monopole cannot wrap a cycle of the CY$_{3}$. 
In this section, we assume a large volume (complex structure) limit and ignore
 any quantum correction.  We do not consider instabilities of wrapped branes 
and assume the supersymmetric cycle condition well behaves. 
To obtain the four-dimensional Einstein frame with a CY$_{3}$ compactification 
we redefine $\beta(t)$ as  
\begin{equation}
\beta(t)=\phi(t)-\frac{1}{2} \ln \Biggl( \frac{V_{\text{CY}_{3}}(t)}{(2\pi \sqrt{\alpha'})^{6}} \Biggr) \label{cy1}
\end{equation}
where we have defined $V_{\text{CY}_{3}}(t) \equiv \int d^{6}y \sqrt{g_{\text{CY}_{3}}(t)}$. The volume of the CY$_{3}$ 
has a time dependence through moduli fields.

If D$p$-branes ($p=0,\,2,\,4,\,6$) are  included minimally in the CY$_{3}$ or its cycle, 
we can represent winding modes of the D$p$-branes as  
\begin{equation}
\begin{split}
m_{0}&=(2\pi \sqrt{\alpha'})^{3}T_{0}  V^{-1/2}_{\text{CY}_{3}},  \\
m^{(A)}_{2}&=(2\pi \sqrt{\alpha'})^{3}T_{2}V^{-1/2}_{\text{CY}_{3}}\times \int_{\varSigma^{A}_{2}} J,    \\
m_{4,\,(A)}&=(2\pi \sqrt{\alpha'})^{3}T_{4} V^{-1/2}_{\text{CY}_{3}} \times \frac{1}{2}\int_{\varSigma_{4,\,A}} J\wedge J,   \\
m_{6}&=(2\pi \sqrt{\alpha'})^{3}T_{6} V^{1/2}_{\text{CY}_{3}},  
\end{split}  \label{cy2}
\end{equation}
where $J$ is the K$\ddot{\text{a}}$hler form,  $\varSigma^{A}_{2}$ and $\varSigma_{4,\,A}$ 
are dual to harmonic form $\omega_{A}$ and $\tilde{\omega}^{A}$. 
The properties of the K$\ddot{\text{a}}$hler form are explained in Appendix \ref{appendixD}. 
The Poincar$\acute{ \text{e}}$ dual gives rise to 
\begin{equation}
\begin{split}
\int_{\varSigma^{A}_{2}} J&=\int_{\text{CY}_{3}}J\wedge \tilde{\omega}^{A}=v^{A},  \\
\int_{\varSigma_{4,\,A}} J \wedge J &=\int_{\text{CY}_{3}}J \wedge J\wedge \omega_{A}\equiv \mathcal{K}_{ABC} v^{B}v^{C}. 
\end{split} \notag 
\end{equation}
Using above equations and (\ref{appD2}), we obtain the following relations between masses of wrapped branes: 
\begin{equation}
\begin{split}
m_{0} \times m_{6}=\Biggl( \frac{1}{\sqrt{\alpha'}} \Biggr)^{2}, \\
\sum_{A \in  h^{1,1}}m^{(A)}_{2} \times  m_{4,\,(A)}=3 \Biggl( \frac{1}{\sqrt{\alpha'}} \Biggr)^{2}. 
\end{split} \label{dual3}
\end{equation}
Eq.(\ref{dual3}) implies that $m_{0}$ ($m^{(A)}_{2}$) is not independent of $m_{6}$ ($m_{4,\,(A)}$). 
In fact, by (\ref{appD4}) and (\ref{appD5}), 
there are relations between those masses as follows: 
\begin{equation}
\begin{split}
m_{6}&=-(2\pi \sqrt{\alpha'})^{-6}\text{Im}\mathcal{N}_{00}m_{0}, \\
m_{4,\,(A)}&=-(2\pi \sqrt{\alpha'})^{-2}\text{Im}\mathcal{N}_{AB}m^{(B)}_{2}.  
\end{split}  \label{dualmass1}
\end{equation}

Note that the dual relation for D$0$- and D$6$-branes is satisfied quite naturally, 
since the dual relation does not require details of the CY$_{3}$. 
For instance, a large CY$_{3}$ with a condition as 
$V_{\text{CY}_{3}} \sim (\mathcal{O}(10^{5})\times 2\pi \sqrt{\alpha'} )^{6}$ gives
 $m_{0} \sim \mathcal{O}(10)$ TeV. 
We consider a large volume by a constant scaling 
\begin{equation}
v^{A}\rightarrow \exp(2s)v^{A}  \label{scaling_1}
\end{equation}
 for all cycles. 
The $m_{2}^{(A)}$ and $m_{4,\,(A)}$ have  the following scalings:
\begin{equation}
\begin{split}
m_{2}^{(A)}&\rightarrow (2\pi \sqrt{\alpha'})^{3}T_{2}V^{-1/2}_{\text{CY}_{3}}\times e^{-s} v^{A}, \\
m_{4,\,(A)}&\rightarrow (2\pi \sqrt{\alpha'})^{3}T_{4}V^{-1/2}_{\text{CY}_{3}}\times e^{+s} \mathcal{K}_{ABC}v^{B}v^{C}
\end{split}  \label{mass_scaling_1}
\end{equation} 
where  $V_{\text{CY}_{3}}\rightarrow \exp(+6s)V_{\text{CY}_{3}}$.
Then light D$2$-branes and heavy D$4$-branes are realized for a positive constant $s$. 
If $\exp(2s)v^{A}$ is chosen as $\mathcal{O}(10^{10}) \times (2\pi \sqrt{\alpha'})^{2}$, 
The $m_{2}^{(A)}$ is of order $\mathcal{O}(10^{15})$ GeV. 
Those results have  same scalings given by (\ref{en1}) and (\ref{mass_torus_scaling}).

The behavior of the D$2$- and D$4$-branes is more complicated than the D$0$-D$6$ case,  
because the dual relation for D$2$- and D$4$-branes cannot restrict all degrees of freedom of moduli fields. 
If we choose a scaling of $v^{A}\rightarrow \exp(2s_{A})v^{A}$, 
the masses of D$2$- and D$4$-branes  vary correspondingly. 
Using (\ref{dual3}) and (\ref{dualmass1}), tuning the scale of cycles of the CY$_{3}$, we can find models having light and heavy  D$2$- and D$4$-branes.

Quite similarly, we can consider a D$3$-brane wrapping over a three-cycle. 
To derive the volume of the 3-cycle from CY$_{3}$ data 
we will use a property of the special Lagrangian submanifold (supersymmetric cycle) \cite{bio_BBS, bio_J, bio_J2}. 
Here, we assume that the condition of the special Lagrangian submanifold well behaves.

In general we call $N$ a special Lagrangian with a constant phase $\exp(i\theta)$, if \cite{bio_BBS, bio_J2}
\begin{equation}
\begin{split}
&J|_{N}=0, \\
&[\sin \theta \text{Re}\Omega-\cos \theta \text{Im}\Omega]|_{N}=0, \\ 
&[\cos \theta \text{Re}\Omega+\sin \theta \text{Im}\Omega]|_{N}=d\text{Vol}_{N} 
\end{split}  \label{csl}
\end{equation}
where the holomorphic three-form is given by 
\begin{equation}
\Omega =Z^{\hat{K}}\alpha_{\hat{K}}-\mathcal{F}_{\hat{K}}\beta^{\hat{K}}, \quad 
 J \wedge J \wedge J=\frac{3i}{4} \Omega \wedge \overline{\Omega}. \label{dual3-2}
\end{equation}
($\alpha_{\hat{K}},\,\beta^{\hat{K}}$) is the dual cohomology basis of ($A^{\hat{K}},\,B_{\hat{K}}$) and 
the real basis of $H^{3}(\text{CY}_{3})$ in that they satisfy 
$\int_{\text{CY}_{3}} \alpha_{\hat{K}} \wedge \beta^{\hat{L}}=\delta^{\hat{L}}_{\hat{K}}$ 
with all other intersections vanishing. 

We will consider  D$3$-brane wrapping over a basis 
of the 3-cycle $A^{\hat{K}}$ or $B_{\hat{K}}$ ($\hat{K}=0,\,1,\,\cdots ,\,h^{2,\,1}$)
as the special Lagrangian submanifold. 
Taking into account (\ref{csl}), we assume the existence of special Lagrangian cycles defined by the following conditions: 
\begin{align}
&\text{Re}\Omega|_{A^{\hat{K}}}=d\text{Vol}_{A^{\hat{K}}}, \quad \text{Im}\Omega|_{A^{\hat{K}}}=0, 
\quad (\theta_{A^{\hat{K}}}=0) \label{sl1}  \\
-&\text{Im}\Omega|_{B_{\hat{K}}}=d\text{Vol}_{B_{\hat{K}}}, \quad \text{Re}\Omega|_{B_{\hat{K}}}=0.  
\quad (\theta_{B_{\hat{K}}}=-\frac{\pi}{2}) \label{sl2}
\end{align}
Those conditions imply 
\begin{equation}
\Omega=\text{Re}Z^{\hat{K}}\alpha_{\hat{K}}-i \text{Im}\mathcal{F}_{\hat{L}}\beta^{\hat{L}} \label{ssc2}
\end{equation}
and 
\begin{equation}
\text{Im}Z^{\hat{K}}=0, \quad \text{Re}\mathcal{F}_{\hat{K}}=0.  \label{ssc3}
\end{equation}
Eq.(\ref{ssc3}) and  Eq.(\ref{app16}) give rise to $\text{Re}\mathcal{M}_{\hat{K}\hat{L}}=0$ and then the Hodge dual of 
($\alpha_{\hat{K}},\,\beta^{\hat{K}}$) defined by (\ref{app13}) and (\ref{app15}) is 
\begin{equation}
\begin{split}
*\alpha_{\hat{K}}&=-(\text{Im}\mathcal{M})_{\hat{K}\hat{L}}\beta^{\hat{L}}, \\
\quad 
*\beta^{\hat{K}}&=(\text{Im}\mathcal{M})^{-1\hat{K}\hat{L}}\alpha_{\hat{L}}.  
\end{split}   \label{ssc4}
\end{equation}
In general, a Hodge dual of a basis includes a linear combination of $\alpha_{\hat{K}}$ and $\beta^{\hat{K}}$ 
as (\ref{app13}). On the other hand, Eq.(\ref{ssc4}) indicates that ($\alpha_{\hat{K}},\,\beta^{\hat{K}}$) is 
decomposed with respect to the Hodge dual, if D$3$-branes can wrap around $A^{\hat{K}}$ and $B_{\hat{K}}$ 
as special Lagrangian cycles with the conditions (\ref{sl1}) and (\ref{sl2}). 
 By (\ref{ssc2}), (\ref{ssc4}) and (\ref{app17}) 
$*\text{Re}\Omega|_{A^{\hat{K}}}=\text{Im}\Omega|_{B_{\hat{K}}}$ is realized and then (\ref{sl1}) and (\ref{sl2}) give rise 
to  $*d\text{Vol}_{A^{\hat{K}}}=-d\text{Vol}_{B_{\hat{K}}}$.

The world volume action of the wrapped D$3$-brane is described by
\begin{equation} 
\begin{split}
S_{\text{D}3}&=-T_{3}\int_{R \times A^{\hat{K}}} d^{3+1}\xi e^{-\phi} \sqrt{-\gamma} \\
&=-\int dt e^{n}\times 
(2\pi\sqrt{\alpha'})^{3} T_{3} V_{\text{CY}_{3}}^{-1/2} \text{Re}Z^{\hat{K}}
 \end{split} \label{dual4}
\end{equation}
where 
\begin{equation}
\text{Vol}_{A^{\hat{K}}}= 
\int_{A^{\hat{K}}}\text{Re}\Omega=
\int_{\text{CY}_{3}} \text{Re}\Omega \wedge \beta^{\hat{K}}=\text{Re}Z^{\hat{K}}.  \notag 
\end{equation} 
Then the mass is given by 
\begin{equation}
m^{(\hat{K})}_{3}=(2\pi\sqrt{\alpha'})^{3} T_{3} V_{\text{CY}_{3}}^{-1/2} \text{Re}Z^{\hat{K}}.  \label{mass_D3}
\end{equation}
The dual volume is also defined by 
\begin{equation}
\text{Vol}_{B_{\hat{K}}}=\int_{B_{\hat{K}}}(-\text{Im}\Omega)
=-\int_{\text{CY}_{3}} \text{Im}\Omega \wedge \alpha_{\hat{K}}=-\text{Im}\mathcal{F}_{\hat{K}}   \notag 
\end{equation}
 and the dual mass is 
\begin{equation}
m_{3,\,(\hat{K})}=-(2\pi\sqrt{\alpha'})^{3} T_{3} V_{\text{CY}_{3}}^{-1/2} \text{Im}\mathcal{F}_{\hat{K}}.  \label{dual5}
\end{equation}
Using (\ref{ssc2}) and the following relation
\begin{equation}
\begin{split}
4V_{\text{CY}_{3}}&=\int_{\text{CY}_{3}} \text{Re}\Omega \wedge \text{Im} \Omega \\
&=-\text{Re}Z^{\hat{K}}\text{Im}\mathcal{F}_{\hat{K}},  
\end{split} \label{dual5_2}
\end{equation} 
$m^{(\hat{K})}_{3}$ and $m_{3,\,(\hat{K})}$ satisfy 
\begin{equation}
\sum_{\hat{K}\in h^{2,\,1}+1} m^{(\hat{K})}_{3} \times m_{3,\,(\hat{K})}=4 \Biggl( \frac{1}{\sqrt{\alpha'}} \Biggr)^{2}.   \label{dual6}
\end{equation}
Those masses also have the following relation: 
\begin{equation}
m_{3,\,(\hat{K})}=-\text{Im}\mathcal{M}_{\hat{K}\hat{L}}m^{(\hat{L})}_{3}. \label{dual7}
\end{equation}

For $h^{1,\,2}+1>1$ the mass of the D$3$-brane 
can take various values. For example, we consider $Z^{1}\rightarrow 0$ and then 
massless D$3$-branes appear on the four-dimensional space-time \cite{bio_ST}. 
If $\text{Re}Z^{\hat{K}}$ has a scaling as $\exp(3s)\text{Re}Z^{\hat{K}}$ for a constant $s$, 
$\mathcal{F}_{\hat{K}}$  also has a scaling as $\exp(3s)\mathcal{F}_{\hat{K}}$, 
because $\mathcal{F}$ is the holomorphic function with the degree two.  
Then, by (\ref{mass_D3}), (\ref{dual5}) and (\ref{dual5_2}) it is found that the mass is of order of the Planck mass 
for a large volume given by $\exp(3s)\text{Re}Z^{\hat{K}}\sim (\mathcal{O}(10^{5}) \times 2\pi \sqrt{\alpha'})^{3}$. 
This result corresponds to (\ref{mass_torus_scaling}).

 In Sec. IV and Sec. V, we have considered the scaling behavior of the mass of wrapped branes.
For the string frame, a large volume of the compactified space and a weak string coupling give rise to heavy branes, 
while, in the four-dimensional Einstein frame, various light particles of the wrapped branes arise after the $T^{6}$ and the CY$_{3}$ 
compactification. 
Those light particles are given by the string scale which is of order the Planck length. 
For example, it has been shown that the mass of a D$0$-brane is of order $\mathcal{O}(10)$ TeV for 
$V_{T^{6}\text{ or CY}_{3}} \sim (\mathcal{O}(10^5) \times 2 \pi \sqrt{\alpha'})^{6}$.   
Therefore, the four-dimensional Einstein frame is quite nontrivial. 
If the light particles  have a weak RR interaction, 
the wrapped branes may become a component of the dark matter. 
We will discuss the RR charge in the next section.

\section{RR charge in the four-dimensional Einstein frame}

In the section IV and V, we have shown that various D$p$-branes become light in the four-dimensional Einstein frame.      
The light D$p$-brane locally interact through the RR flux. 
We have to estimate th RR charge, because the dark matter should have a very weak interaction.   
To estimate the coupling of the RR flux, 
we will analyze the RR flux and the Wess-Zumino (WZ) term by the four-dimensional Einstein frame. 
If D$p$-brane has a weak coupling, the light D$p$-brane is a dark matter candidate. 
We do not consider a contribution of the D-instanton ($p=-1$) potential, for simplicity. 
In this section, the RR potential is a function of the four-dimensional coordinates, 
while the fields on the D$p$-brane and the scale factor depend only on the time coordinate.

The action of the RR flux and the WZ term of a D$p$-brane wrapping over a ($m_{1}\cdots m_{p}$)-cycle is given by \cite{bio_P} 
\begin{align}
S_{\text{RR}}^{(m_{1}\cdots m_{p})}&=-\frac{1}{4\kappa_{10}^{2}} \int_{M^{10}}  d^{10}X \sqrt{-G}|F_{p+2}|^{2}, \label{e40} \\
S_{\text{WZ}}^{(m_{1}\cdots m_{p})}&=\mu_{p}\int_{R\times \varSigma_{p}} \exp( 2\pi \alpha' F_{(2)} ) \wedge \sum_{q}C_{q} \label{e41}
\end{align}
where 
\begin{equation}
|F_{p+2}|^{2} \equiv  \frac{1}{(p+2)!}F_{\mu_{1} \cdots \mu_{p+2}}F^{\mu_{1} \cdots \mu_{p+2}}, \label{e42}
\end{equation}
$F_{p+2}=dC_{p+1}$ is the RR flux and $F_{(2)}$ is the gauge field on the D-brane. 
In the string frame the RR charge has a relation given by \cite{bio_P}
\begin{equation}
\mu_{p}^{2}=T_{p}^{2}.  \label{e43}
\end{equation}

We will consider the dimensional reduction of the WZ term. 
The gauge field on the D$p$-brane is given by 
\begin{equation}
F_{(2)}=\dot{A}_{a}(t) dt \wedge d \xi^{a}.  \label{e44}
\end{equation}
This equation leads to a relation  $F_{(2)} \wedge \cdots \wedge F_{(2)}=0$. 
The RR potential of a D$p$-brane wrapping the specific ($m_{1}\cdots m_{p}$)-cycle  can be expanded as 
\begin{align}
&C_{p+1}=C^{(m_{1}\cdots m_{p})}_{\mu}(x) \frac{dX^{\mu}}{dt}dt \wedge d\xi^{m_{1}} \wedge \cdots \wedge d\xi^{m_{p}} \notag \\ 
&  +\sum_{b=p+1}^{6} C^{(m_{1}\cdots m_{p})}_{m_{b }}(x) 
\frac{dX^{m_{b }}}{dt}dt \wedge d\xi^{m_{1}} \wedge \cdots \wedge d\xi^{m_{p}} \label{e46}
\end{align}
where $X^{i}$ and $X^{m_{b }}$ are transverse coordinates of the wrapped D$p$-brane. 
After the Kaluza-Klein reduction, $C^{(m_{1}\cdots m_{p})}_{m_{b }}(x)$ is a 
scalar field on the four-dimensional space-time and
the WZ term is  
\begin{align}
&S^{(m_{1}\cdots m_{p})}_{\text{WZ}}=\int_{M_{4}}d^{4}x (2\pi \sqrt{\alpha'})^{p}  \mu_{p} \delta^{3}({\bf x}-{\bf X}(t))  \notag \\
&\times  \Biggl( C^{(m_{1}\cdots m_{p})}_{\mu}(x) \frac{dX^{\mu}}{dt}
+\sum_{b=p+1}^{6}C^{(m_{1}\cdots m_{p})}_{m_{b }}(x) \frac{dX^{m_{b }}}{dt} 
\Biggr)  .  \label{es1} 
\end{align}

First, we will consider the $T^{6}$ compactification. 
We perform the Kaluza-Klein reduction of the kinetic term of the RR flux and  expand the RR potential 
on the ten-dimensional space-time as 
$C_{p+1}=( C^{(m_{1}\cdots m_{p})}_{\mu}dx^{\mu}+\sum_{b=p+1}^{6} 
C^{(m_{1}\cdots m_{p})}_{m_{b }} dy^{m_{b }} ) 
\wedge dy^{m_{1}} \wedge \cdots \wedge dy^{m_{p}}$. 
Substituting this equation for (\ref{e40}) and using (\ref{e3}), (\ref{e5}), (\ref{e_newton}), (\ref{e7}) and (\ref{e43}), 
we obtain the following effective action of the RR flux in the four-dimensional Einstein frame:   
\begin{align}
&S_{\text{RR}}^{(m_{1}\cdots m_{p})}= \int_{M^{4}}  d^{4}x \sqrt{-g_{4}} \notag \\
&\times \Biggl( -\frac{1}{4 ( g^{(m_{1}\cdots m_{p})}_{p} )^{2}} \tilde{F}^{(m_{1}\cdots m_{p})}_{\mu\nu}
\tilde{F}^{(m_{1}\cdots m_{p})\mu\nu} \notag \\ 
& - \sum_{b=p+1}^{6}\frac{1}{ 2 ( \tilde{g}^{(m_{1}\cdots m_{p},\,m_{b })}_{p}  )^{2} } \partial_{\mu} 
\tilde{C}^{(m_{1}\cdots m_{p})}_{m_{b }} \partial^{\mu} 
\tilde{C}^{(m_{1}\cdots m_{p})}_{m_{b }} \Biggr),  \label{e47} \\
&( g^{(m_{1}\cdots m_{p})}_{p}   )^{2}\equiv 2 \pi \exp(-\overline{\lambda}+2\sum_{a=1}^{p} \lambda_{m_{a}})  , \label{e47_2} \\
& ( \tilde{g}^{(m_{1}\cdots m_{p},\,m_{b})}_{p}  )^{2} \equiv 2 \pi \exp(-2\phi+2\sum_{a=1}^{p} \lambda_{m_{a}}+2\lambda_{m_{b }}) \label{e47_3}
\end{align}
where 
\begin{equation}
\begin{split}
&\tilde{F}^{(m_{1}\cdots m_{p})}_{\mu\nu}=\partial_{\mu} \tilde{C}^{(m_{1}\cdots m_{p})}_{\nu}-\partial_{\nu} 
\tilde{C}^{(m_{1}\cdots m_{p})}_{\mu},  \\
&\tilde{C}^{(m_{1}\cdots m_{p})}_{\mu} \equiv (2\pi \sqrt{\alpha'})^{p}\mu_{p}C^{(m_{1}\cdots m_{p})}_{\mu}, \\
&\tilde{C}^{(m_{1}\cdots m_{p})}_{m_{b }} \equiv (2\pi \sqrt{\alpha'})^{p}\mu_{p}C^{(m_{1}\cdots m_{p})}_{m_{b }}.
\end{split} \label{rescales}
\end{equation} 
The effective action of $\tilde{F}_{\mu\nu}^{(m_{1}\cdots m_{p})}$ has no dependence of the dilaton. 
This is the fact on $N=2$ supergravity \cite{bio_DLP}. The dilaton field lives in a hypermultiplet  and does not  
couple with a vector multiplet.

We find that $g^{(m_{1}\cdots m_{p})}_{p}$ satisfies the electric-magnetic dual relation: 
\begin{equation}
g^{(m_{1}\cdots m_{p})}_{p} \times g^{(m_{p+1}\cdots m_{6})}_{6-p}=2 \pi .  \label{charge_dual1}
\end{equation}
By (\ref{charge_dual1}) the coupling $g^{(m_{1}\cdots m_{p})}_{p}$ can take a small value, 
while the dual coupling $g^{(m_{p+1}\cdots m_{6})}_{6-p}$ becomes large. 
The dual relation (\ref{charge_dual1}) is also understood as follows. 
If we add a trivial quantity which is proportional to $\int F_{p+2} \wedge G_{8-p}$, $G_{8-p}=d D_{7-p}$ to the action (\ref{e40}) 
and eliminate 
$F_{p+2}$, we obtain $\int G_{8-p} \wedge * G_{8-p}$ by the dual field $G_{8-p}=*F_{p+2}$. 
Performing the Kaluza-Klein reduction, we can take the coupling of the dual field, $g^{2}_{6-p}$ in term of the 
definition (\ref{e47_2}).  
We consider $\tilde{g}^{(m_{1}\cdots m_{p},\,m_{b})}_{p}$ later.

Secondly, we would like to discuss the CY$_{3}$ compactification. 
The property considered above is general and  
it is expected that the charge relations can be satisfied in the cases of the CY$_{3}$ compactification.  
However, in the CY$_{3}$ compactification, the square of the coupling becomes 
a matrix defined by the K$\ddot{\text{a}}$hler or the complex structure moduli after the Kaluza-Klein reduction 
\cite{bio_BGHL, bio_LM, bio_GL, bio_GL2, bio_DLP}.  
For example, we would like to consider the type IIA supergravity compactified on a CY$_{3}$. 
After rescaling as 
\begin{equation}
C_{p+1} = \mu^{-1}_{p} \tilde{C}_{p+1}, \notag 
\end{equation} 
the RR flux is given by 
\begin{equation}
F_{2}=\mu_{0}^{-1}\tilde{F}, \qquad
F_{4}=\mu_{2}^{-1}\tilde{F}^{A}\omega_{A}.   \notag 
\end{equation}
Then the gauge kinetic term of the vector multiplets is given by 
\begin{equation}
\begin{split}
&-\frac{1}{2\kappa_{10}^{2}}  \int_{M_{10}} \frac{1}{2} (F_{2}\wedge *F_{2}+F_{4} \wedge * F_{4}) \\
=&  \int_{M_{4}} -\frac{1}{2} \Biggl[ 
\Biggl\{ \frac{-1}{2\pi} \frac{\text{Im}\mathcal{N}_{00}}{(2\pi \sqrt{\alpha'})^{6}} \Biggr\}\tilde{F} \wedge * \tilde{F} \\
& \qquad \qquad +\Biggl\{ \frac{-1}{2\pi} \frac{\text{Im}\mathcal{N}_{AB}}{(2\pi \sqrt{\alpha'})^{2}} \Biggr\} \tilde{F}^{A} \wedge * \tilde{F}^{B} \Biggr]
\end{split}  \label{dualfield1}
\end{equation}
where ($A=\,1,\,\cdots ,\,h^{1,1}$). 
$\text{Im}\mathcal{N}_{00}$ and $\text{Im}\mathcal{N}_{AB}$ are defined by (\ref{appD5}). 
Introducing a dual field $\tilde{G}_{6}=\tilde{F}^{\text{dual}}_{A}\tilde{\omega}^{A}$ and 
$\tilde{G}_{8}=\tilde{F}^{\text{dual}}d\text{Vol}_{\text{CY}_{3}}$ 
which are related to the flux of D$4$- and D$6$-brane , we will consider the following term:  
\begin{equation}
\begin{split}
\frac{1}{2\pi}\int_{M_{10}}\tilde{F}_{2}\wedge \tilde{G}_{8}&=\frac{1}{2\pi} \int_{M_{4}} V_{\text{CY}_{3}}\tilde{F} \wedge \tilde{F}^{\text{dual}},  \\
\frac{1}{2\pi}\int_{M_{10}}\tilde{F}_{4}\wedge \tilde{G}_{6}&=\frac{1}{2\pi}\int_{M_{4}} \tilde{F}^{A} \wedge \tilde{F}_{A}^{\text{dual}}
\end{split}  \label{dualfield2}
\end{equation}
where we have used (\ref{appD7}). Adding (\ref{dualfield2}) to (\ref{dualfield1}) and 
integrating out $\tilde{F}$ and $\tilde{F}^{A}$ ($A=1,\,2,\,\cdots ,\,h^{1,1}$), the gauge kinetic term of (\ref{dualfield1}) is mapped to 
\begin{equation}
\begin{split}
\int_{M_{4}} -\frac{1}{2} \Biggl[ \Biggl\{ \frac{-1}{2\pi} (2\pi \sqrt{\alpha'})^{6}(\text{Im}\mathcal{N}^{-1})^{00} \Biggr\}
\tilde{F}^{\text{dual}} \wedge * \tilde{F}^{\text{dual}} \\ 
+\Biggl\{ \frac{-1}{2\pi} (2\pi \sqrt{\alpha'})^{2}(\text{Im}\mathcal{N}^{-1})^{AB} \Biggr\} 
\tilde{F}^{\text{dual}}_{A} \wedge * \tilde{F}^{\text{dual}}_{B}  \Biggr]   \label{dualfield3}
\end{split}
\end{equation} 
where the inverse matrix on the couplings is defined by (\ref{appD6}) and 
$*^{2} \tilde{F}^{\text{dual}}_{\hat{A}}=- \tilde{F}^{\text{dual}}_{\hat{A}}$  has been used  for a four-dimensional Lorentz manifold. 
We can read off the gauge couplings is as follows:
\begin{equation}
\begin{split}
(g^{-2})_{00}&=\frac{-1}{2\pi} \frac{\text{Im}\mathcal{N}_{00}}{(2\pi \sqrt{\alpha'})^{6}}, \\
(g^{-2})_{AB}&=\frac{-1}{2\pi} \frac{\text{Im}\mathcal{N}_{AB}}{(2\pi \sqrt{\alpha'})^{2}}, \\
(g^{-2}_{\text{dual}})^{00}&= \frac{-1}{2\pi} (2\pi \sqrt{\alpha'})^{6}(\text{Im}\mathcal{N}^{-1})^{00}, \\
(g^{-2}_{\text{dual}})^{AB}&= \frac{-1}{2\pi} (2\pi \sqrt{\alpha'})^{2}(\text{Im}\mathcal{N}^{-1})^{AB}. 
\end{split}  \label{dualcouplings1}
\end{equation}
Those couplings satisfy 
\begin{equation}
(g^{-2})_{\hat{A}\hat{B}}(g^{-2}_{\text{dual}})^{\hat{B}\hat{C}}= (2\pi)^{-2}\delta^{\hat{C}}_{~\hat{A}}  \label{inverse_dual_charge}
\end{equation}
where ($\hat{A}=0,\,1,\,2,\, \cdots ,\,h^{1,1} $). 
The above relation  has the same structure to  (\ref{charge_dual1}). 
In type IIB theory, the coupling of the flux of a D$3$-brane wrapping over a three-cycle 
is given by replacing $\mathcal{N}$ with $\mathcal{M}$ defined by (\ref{app16}) \cite{bio_GL2} and 
taking a suitable normalization constant. 

We will consider the scaling behavior of $g^{(m_{1}\cdots m_{p})}_{p}$ and $\tilde{g}^{(m_{1}\cdots m_{p},\,m_{b })}_{p}$, 
assuming a constant dilaton and constant moduli fields. 
We should notice that $\tilde{g}^{(m_{1}\cdots m_{p},\,m_{b})}_{p}$ is not the true coupling  of $C^{(m_{1}\cdots m_{p})}_{m_{b }}$, 
since the kinetic term of $X^{m_{b }}(t)$ has a coupling with moduli fields such as
$e^{-2\beta+2\lambda'} (d X^{m_{b }}/dt)^{2}$
 in  (\ref{e9}). 
To normalized the kinetic term, we define the following equations:
\begin{equation}
\begin{split}
&\frac{d X^{m_{b }}}{dt}\equiv e^{\beta-\lambda_{m_{b}}} \frac{d \tilde{X}^{m_{b }}}{dt}, \\
&\tilde{C}^{(m_{1}\cdots m_{p})}_{m_{b }}\equiv e^{-\beta+\lambda_{m_{b}}} \Psi ^{(m_{1}\cdots m_{p})}_{m_{b }}. 
\end{split}
\end{equation}
Then, the kinetic term is proportional to $e^{-2n}(d \tilde{X}^{m_{b }}/dt)^{2}$ in (\ref{e8}). 
Using the above relations, the kinetic term of $\Psi ^{(m_{1}\cdots m_{p})}_{m_{b }}$ is given by
\begin{equation} 
 \sum_{b=p+1}^{6} \frac{-1}{2(g^{(m_{1}\cdots m_{p},\,m_{b })}_{p})^{2}} \partial_{\mu} 
\Psi ^{(m_{1}\cdots m_{p})}_{m_{b }} \partial^{\mu} 
\Psi ^{(m_{1}\cdots m_{p})}_{m_{b }}
\end{equation}
 where  the coupling is correspond with (\ref{e47_2}): 
\begin{align}
(g^{(m_{1}\cdots m_{p})}_{p})^{2}=(g^{(m_{1}\cdots m_{p},\,m_{b })}_{p})^{2}. \label{n1}
\end{align}

We will consider the isotropic case such as $\lambda_{m}=\lambda_{n}$. 
$g^{(m_{1}\cdots m_{p})}_{p}$ and $g^{(m_{1}\cdots m_{p},\,m_{b })}_{p}$ 
can take a small value for $p\leq 2$ and $e^{\lambda'}\gg 1$. 
We will take $e^{\lambda'} \sim \mathcal{O}(10^{5})$ for instance. 
The coupling is given by
\begin{equation}
(g^{(m_{1}\cdots m_{p})}_{p})^{2}=(g^{(m_{1}\cdots m_{p},\,m_{b })}_{p})^{2} \sim \mathcal{O}(10^{10(p-3)}).  \label{order_RR_coupling}
\end{equation}
According to the relation, 
$\text{D}0$-brane has the very weak interaction $g^{2}_{(0)} \sim \mathcal{O}(10^{-30})$. 
For wrapped D$1$- and D$2$-brane 
the coupling  is given by $(g^{(m_{1})}_{1})^{2} \sim \mathcal{O}(10^{-20})$ 
and $(g^{(m_{1}m_{2})}_{2})^{2} \sim \mathcal{O}(10^{-10})$, respectively. 
The coupling of D$3$-brane has $g^{(m_{1}m_{2}m_{3})}_{3} \sim \mathcal{O}(1)$. 
In the anisotropic case, $\lambda_{m} \neq \lambda_{n}$,  
Eq.(\ref{e47_2}) with (\ref{charge_dual1}) has the same classification on the mass of D-branes in section IV .

We also investigate scaling of the charges, controlling scales of the Calabi-Yau by hand.  
In the case with a CY$_{3}$ compactification small couplings of wrapped D$0$- and D$2$-branes 
are realized by a large $-\text{Im}\mathcal{N}_{\hat{A}\hat{B}}$, 
because of $-\text{Im}\mathcal{N}_{\hat{A}\hat{B}} \propto (g^{-2})_{\hat{A}\hat{B}}$. 
For example, using (\ref{appD4}) and (\ref{appD5}), 
we consider a large volume of the CY$_{3}$ 
by a scaling, $v^{A}\rightarrow \exp(2s)v^{A}$ for a positive constant $s$.  
This condition gives rise to 
\begin{equation}
\begin{split}
\text{Im}\mathcal{N}_{00}&\rightarrow \exp(6s) \text{Im}\mathcal{N}_{00}, \\
\text{Im}\mathcal{N}_{AB}&\rightarrow \exp(2s) \text{Im}\mathcal{N}_{AB}. 
\end{split}  \label{CY_coupling_2}
\end{equation}
Using (\ref{dualcouplings1}) and (\ref{appD5}), the coupling of the D$0$-brane can be $g^{2}\sim \mathcal{O}(10^{-30})$
for $V_{\text{CY}_{3}}\sim (\mathcal{O}(10^{5})\times 2\pi \sqrt{\alpha'})^{6}$. 
Then the weak coupling is realized for wrapped D$0$- and D$2$-branes. 
The coupling of D$3$-branes which is related with the matrix $\mathcal{M}_{\hat{K}\hat{L}}$ in (\ref{app16}) 
has different behavior from the flux of type IIA. 
The coupling involves the holomorphic function with the degree two, 
$\mathcal{F}(\exp(3s')Z^{\hat{K}})=\exp(6s')\mathcal{F}(Z^{\hat{K}})$ by $Z^{\hat{K}}\rightarrow \exp(3s')Z^{\hat{K}}$.    
Then, (\ref{app16}) implies that the coupling matrix $\mathcal{M}$ does not change for the scaling of $Z^{\hat{K}}$. 
We may require details of the topological data of the CY$_{3}$ to analyze the scaling of $\mathcal{M}$, 
however we do not consider the details in this paper.

\section{fluctuations of moduli fields}

In previous sections, we have investigated the mass and the RR charge of the wrapped branes to consider 
a possibility of a dark matter candidate. 
It has been shown that there are cases where the wrapped D$p$-brane has a light mass and a weak RR charge in the four-dimensional Einstein frame, 
if the scale of the compactified space satisfies a condition discussed in section IV, V and VI.    
However, we have to discuss an interaction in term of fluctuations of moduli fields, 
because the fluctuations give rise to propagations of scalar interactions between branes. 
In fact, from the ten-dimensional point of view, the radial moduli is one of components of the gravitational fields  
and then the fluctuations of the moduli also give the interaction between the D-branes \cite{bio_P}. 
In this section, we would like to comment on a charge of the interaction without the analysis of the cosmological perturbation. 
We consider the $T^{6}$ compactification only.

The kinetic term of the moduli fields is diagonal in (\ref{e4}). To canonically normalize the kinetic term of the fluctuations, 
we define  
\begin{equation}
\lambda_{m}=\lambda^{(0)}_{m}+\sqrt{8 \pi G_{4}}\,\delta \tilde{\lambda}_{m}  \label{moduli_fluctuation_1}
\end{equation}
 where 
$\lambda^{(0)}_{m}$ is a fixed constant.  
Then, using (\ref{e8}), the source term of the interaction is given by 
\begin{equation}
\begin{split}
\delta S^{(m_{1}\cdots m_{p})}_{\text{D}p}&\propto -(2\pi \sqrt{\alpha'})^{p}T_{p} \sqrt{8 \pi G_{4}} 
 \exp \Bigl( \sum_{a=1}^{p}\lambda^{(0)}_{m_{a}}-\frac{\overline{\lambda}^{(0)}}{2}  \Bigr) \\
&\times \int dte^{n(t)} \Biggl( -\frac{1}{2}\delta \overline{\tilde{\lambda}}+\sum_{a=1}^{p} 
\delta \tilde{\lambda}_{m_{a}}  \Biggr). \label{n2}
\end{split}
\end{equation} 
The coefficient of the above equation gives the charge in term of the fluctuation. The magnitude of the interaction is 
given by the square of the coefficient:     
\begin{equation}
\pi \exp \Bigl( 2\sum_{a=1}^{p}\lambda^{(0)}_{m_{a}}-\overline{\lambda}^{(0)}  \Bigr). \label{n3}
\end{equation}
It is found that the scaling of (\ref{n3}) is same to (\ref{e47_2}) and the similar dual relation such as (\ref{charge_dual1}) is satisfied. 
If $2\sum_{a=1}^{p}\lambda^{(0)}_{m_{a}}-\overline{\lambda}^{(0)}<0$ is satisfied, the charge becomes weak. 
A condition, $\lambda^{(0)}_{m}=\lambda^{(0)}_{n}$ also realizes a weak interaction on the fluctuations at a large scale of $T^{6}$ 
for D$0$-, D$1$- and  D$2$-branes.

Taking into account results discussed in previous sections, 
we find that there are cases in which wrapped D$p$-branes have the light mass, the weak RR charge and the weak interaction 
on the fluctuations of the moduli fields in the four-dimensional Einstein frame, considering a large volume of the compactified space. 
For instance, the mass  of a D$0$-brane is of the order of  $\mathcal{O}(10)$ TeV for 
$V_{T^{6}\text{ or CY}_{3}} \sim (\mathcal{O}(10^5) \times 2 \pi \sqrt{\alpha'})^{6}$. 
The square of the charge on the RR flux and the fluctuation of the moduli fields is also of order $\mathcal{O}(10^{-30})$. 
Therefore, a possibility of the dark matter arises from the wrapped D$p$-branes in the four-dimensional Einstein frame.

\section{ D$1$-KK$5$ brane gas system}

We have considered the behavior of masses and charges of wrapped branes in the four-dimensional 
Einstein frame, adjusting the compactification scale by hand. 
However, it is not necessarily possible to tune the scale of the compactification freely, 
because this tuning should be consistent with a weak string coupling to consider 
the effective field theory at a low energy after the moduli stabilization. 
In this section,  
we will investigate a moduli stabilization in the $T^{6}$ compactification,  
using the brane gas system of D$1$-branes and KK$5$-monopoles in type IIB string theory \cite{bio_SS2}. 
A condition  for a large volume and a weak string coupling will be explained. We  will estimate the mass of the wrapped branes.  

At a  matter-dominated era, the velocity of wrapped branes may vanish, on average, in a very large scale.  
Then, we will consider the case of 
$\dot{X}^{i}(t)\simeq 0   $
 which means that wrapped brane gases are 
completely non-relativistic particles  \cite{bio_SS2}. 
We have assumed the cancellation of the total charge under the brane gas approximation and then we can use 
results in section III.  
We analyze a type IIB brane gas model in which the D1-branes wrap over each 1-cycle and 
the KK$5$-monopoles wrap over the ($45678$)-cycle and its cyclic permutations. 
We will take the following initial condition,  
\begin{equation}
f_{m_{a}} = 2 \pi \alpha' f, \qquad v^{m_{a}}=0 \label{e22}
\end{equation}
to obtain the analytic value 
of $\lambda_{a}(t)$ and $\beta(t)$ at a minimum. 
The choice of this initial condition represents 
the fact where the initial gauge fields on the D$1$-branes are same for each cycle.
This condition may be natural for the isotropic expansion of the internal space. 
The above condition provides us the following energy density: 
\begin{equation}
U^{\text{IIB}}=e^{-3A(t)} \times ( N_{\text{D}1}m_{1}+ N_{\text{KK}5}m_{\text{KK}5}) \label{e23}
\end{equation}
where $N_{\text{D}1}$ and $N_{\text{KK}5}$ are the number density of the point particles. 
This number density is a constant because the number density is decided at present as $e^{A(t=0)}=1$. 
The scaling of the number density is controlled by the scale factor $e^{A(t)}$.
$m_{\text{D}1}$ and $m_{\text{KK}5}$ are given by  
\begin{align}
&m_{1} \equiv (2 \pi \sqrt{\alpha'})T_{\text{1}}\sum_{a=4}^{9}
e^{-\frac{1}{2}\overline{\lambda}(t)+\lambda_{a}(t)}(1+f^{2}e^{2\beta(t)+\overline{\lambda}})^{\frac{1}{2}}, \label{e24} \\
&m_{\text{KK}5} \equiv (2 \pi \sqrt{\alpha'})^{5}T_{\text{KK$5$}}
\sum_{a=4}^{9} e^{-\beta(t)+\lambda_{a}(t)}.  \label{e25}
\end{align}
If all moduli fields are fixed, 
$U^{\text{IIB}}$ is equal to the energy density of a pressureless matter as $\rho \sim e^{-3A}$.

By (\ref{e24}) and (\ref{e25})  $U^{\text{IIB}}$ takes a positive value. 
Thus the minimum is given by $\partial U^{\text{IIB}} /\partial \beta=\partial U^{\text{IIB}} /\partial\lambda_{a} =0$. 
($\frac{\partial}{\partial \lambda_{a}}-\frac{\partial}{\partial \lambda_{b}})U^{\text{IIB}}=0$ gives
\begin{equation}
e^{\lambda_{a}(t)}=e^{\lambda_{b}(t)}\equiv e^{\lambda'(t)} \label{e26}
\end{equation}
for ($a, b=4,\,5,\,\dots ,\,9$).
Using (\ref{e26}) and the minimum conditions, we obtain  analytic values of moduli fields at the minimum \cite{bio_SS2}: 
\begin{align}
e^{2 \beta_{\text{min.}}}&=\sqrt{\frac{2 }{f^{2}}}\frac{N_{\text{KK}5}}{N_{\text{D}1}}, \label{e27} \\ 
\quad e^{2 \lambda'_{\text{min.}}}&=\Biggl( \frac{ 1}{2f^{2}} \Biggr)^{\frac{1}{6}}
\Biggl( \frac{N_{\text{D}1}}{N_{\text{KK}5}} \Biggr)^{\frac{1}{3}}. \label{e28}
\end{align}
As shown in Sec. III, $f$ is an integer therefore stabilized value of moduli fields cannot continuously connect to
another stabilized values under the $T^{6}$ compactification.   

These equations have a relation  as $e^{2\phi_{\text{min.}}}=e^{2\beta_{\text{min.}}+6\lambda'_{\text{min.}}}=f^{-2}$.  
The dilaton field at the minimum is given by 
\begin{equation}
e^{2 \phi_{\text{min.}}}=\frac{1}{f^{2}} \label{e29}
\end{equation}
where we have used (\ref{e3}), (\ref{e27}) and (\ref{e28}). 
By  (\ref{e28}) and (\ref{e29}) there is a relation between $N_{\text{D}1}$ and $N_{\text{KK}5}$ as follows:  
\begin{equation}
\frac{N_{\text{D}1}}{N_{\text{KK}5}} =\sqrt{2} e^{-\phi_{\text{min.}}+6\lambda'_{\text{min.}}}.  \label{e_number}
\end{equation}
If a large volume of the $T^{6}$, $\exp(\lambda'_{\text{mim.}}) \gg1$ and a weak string coupling, 
\begin{equation}
g_{s}=\exp(\phi_{\text{mim.}})=\sqrt{\frac{1}{f^2}}\ll 1 \label{weak_string_coupling}  
\end{equation}
 are imposed at the minimum, (\ref{e_number}) gives the following condition:
\begin{equation}
N_{\text{D}1} \gg N_{\text{KK}5}.  \label{number_condition}
\end{equation} 
(\ref{number_condition}) indicates that the wrapped D$1$-brane gas dominates the  components.
The number density of KK$5$ has very low number density by the relation (\ref{e_number})
and then  KK$5$-monopoles may survive the annihilation by the heavy mass and the low number density on the three-dimensional space.

Eq.(\ref{e28}) and Eq.(\ref{e29}) show that the string coupling $g_{s}=e^{\phi_{\text{min.}}}$ is related to 
the scale of the compactification through the initial condition of the electric fields on the D$1$-brane. 
We require a weak string coupling $g_{s}\ll 1$ and a large scale of the compactification $e^{\lambda'_{\text{min}}} \gg 1$. 
If we take the specific initial condition as $1 \ll f^{2}$ in (\ref{e29}), 
the weak string coupling is realized. 
However, DBI action has upper bound for the electric fields. 
This bound is given by 
\begin{equation}
1-\mathcal{A}(t)>0 \label{e30}
\end{equation} 
in (\ref{e8}) because of $\dot{X}^{i}(t)\simeq 0$. 
We have to check that the initial condition satisfies (\ref{e30}) at the minimum.
Substituting (\ref{e27}) and (\ref{e28}) for (\ref{e12}), we obtain the following condition 
for a D$1$-brane: 
\begin{equation}
1-\mathcal{A}_{\text{min.}}=
1-\frac{f^{2}e^{2\beta_{\text{min.}}+6\lambda'_{\text{min.}}}}{1+f^{2}e^{2\beta_{\text{min.}}+6\lambda'_{\text{min.}}}}
=\frac{1}{2}. \label{e31}
\end{equation} 
Therefore we can take the initial condition satisfying $1 \ll f^{2}$ because 
(\ref{e31}) does not depend on any initial condition in this model. 
By (\ref{e_number}) and (\ref{e30}) we can take the weak string coupling and the large volume of $T^{6}$ simultaneously.

Finally, we would like to estimate the mass scale of the  branes.  
At the minimum given by (\ref{e28}) and (\ref{e29}), the scale of $T^{6}$ is isotropic, i.e. 
$e^{\lambda_{a\, \text{min.}}}=e^{\lambda'_{\text{min.}}}$ for ($a=4,\, \cdots ,\,9$).
Then, the mass of one of D$1$-branes and KK$5$-monopoles defined by (\ref{e24}) and (\ref{e25}) is given by 
\begin{equation}
\begin{split}
m^{(m_{1})}_{1}|_{\text{min.}}&=\frac{e^{-2\lambda'_{\text{min.}}}}{2}\times m_{\text{Planck}} , \\
m^{(m_{1}\cdots m_{5})}_{\text{KK}5}|_{\text{min.}}&=\frac{e^{-\phi_{\text{min.}}
+4\lambda'_{\text{min.}}}}{2 \sqrt{2}}\times m_{\text{Planck}}
\end{split} \label{e32}
\end{equation}
where $m_{\text{Planck}}\equiv G_{4}^{-1/2}=2\sqrt{2}/\sqrt{\alpha'}\sim \mathcal{O}(10^{19})$ GeV is the four-dimensional Planck mass.  
If we take a weak string coupling $e^{\phi_{\min.}} \ll 1$ and the large scale, $e^{\lambda'_{\text{min.}}} \gg 1$, 
the mass of KK$5$-monopole becomes heavier than the Planck mass.  
On the other hand, D$1$-branes become light particles under the weak string coupling and the large scale of $T^{6}$.  
 If we consider $e^{\lambda'_{\text{min.}}}\sim \mathcal{O}(10^{5})$, 
the order of the mass  is given by
$ m_{1}|_{\text{min.}} \sim \mathcal{O}(10^{9}) \, \text{GeV}$.

In the present section, we have considered the  brane gas model constructed by D$1$-KK$5$ as a simple model of the moduli stabilization. 
It is found that (\ref{weak_string_coupling}) and (\ref{number_condition}) are required to realize a large volume 
and a weak string coupling at a minimum of the potential and the D$1$-brane gas becomes light.

\section{Summary and discussions}

We have considered a possibility of 
the dark matter candidate for wrapped branes in brane gas cosmologies based on the type II string theories.  
Using the four-dimensional Einstein frame, 
we have investigated the mass, the RR charge and 
the interaction on fluctuations of the moduli fields. 
This analysis has been done by the description of the effective field theory, taking the string scale as 
$1/\sqrt{\alpha'} \sim m_{\text{Planck}}$.  
A large volume of the compactified space and a weak string coupling are required 
to suppress string excitations and to obtain the perturbative description.

We have found models where D-branes wrapping over cycles of a compactified space has 
a light mass by the large volume and the weak string coupling, 
after the $T^{6}$ and the CY$_{3}$ compactification, while the mass becomes very heavy in the string frame. 
The masses satisfy the electric-magnetic dual relation between a D$p$- and a D$(6-p)$-brane. 
The four-dimensional Einstein frame gives rise to 
the dual relation which is time-independent although each mass depends on the time coordinate through moduli fields. 
The string and the $(d+1)$-dimensional ($d \neq 3$) Einstein frame, on the other hand, derive the dual relation depending on the 
time variable. Thus, the four-dimensional Einstein frame is the special case and   
the electric-magnetic dual relation gives the mass hierarchy.  
For example, the mass of a D$0$-brane is of order $\mathcal{O}(10)$ TeV, if we take 
a large volume as $V_{T^{6} \text{ or CY}_{3}} \sim (\mathcal{O}(10^{5}) \times 2 \pi \sqrt{\alpha'})^{6}$. 
Similar dual relation which is time-independent is realized for (F$1$, NS$5$) and for ( the momentum of F$1$, KK$5$) 
for the $T^{6}$ compactification. 
The effective charges of the RR flux and of the moduli fluctuation 
have been investigated. 
The charges between the D$p$- and D$(6-p)$-brane also satisfy 
the electric-magnetic dual relations and then there are cases in which the wrapped D$p$-branes obtain 
a small charge for the large volume and the weak string coupling.  
For instance, the square of the charge of a D$0$-brane is of  order   
$\mathcal{O}(10^{-30})$ for  
$V_{T^{6} \text{ or CY}_{3}} \sim (\mathcal{O}(10^{5}) \times 2 \pi \sqrt{\alpha'})^{6}$.

We have considered the behavior of the mass of wrapped branes, using a toy model constructed by a D$1$-KK$5$ 
brane gas \cite{bio_SS2} where radial moduli fields of the $T^{6}$ and the dilaton are simultaneously stabilized. 
The effective field description requires a large volume of the compactified space and a weak string coupling, 
however a point where all moduli fields are stabilized depends on models and 
the realization of the condition is not trivial. 
In this model, the condition of the large volume and the weak string coupling 
imposes (\ref{weak_string_coupling}) and (\ref{number_condition}) at a minimum of the potential of the moduli fields  
and then light D$1$-branes appear for a large volume of the $T^{6}$.

Taking the four-dimensional Einstein frame is very simple, 
however quite non-trivial results among masses and the charges are realized.  
We should consider the light wrapped branes in string cosmologies because of the very small charges. 
Those ingredients may be a candidate for the dark matter. 
We have considered no NSNS two form and no quantum correction. 
Those quantities may give the wrapped branes interesting results, using explicit CY$_{3}$ compactifications.

To check the possibility on the dark matter more rigorously 
we have to investigate the density perturbation of the dust of wrapped branes.  
If domains with dense gases appear on the three-dimensional space, 
 interactions on the RR flux and on the moduli fluctuations may 
have an interesting role for the evolution of the density,   
because of $\mathcal{O}((g^{(m_{1}\cdots m_{p})}_{p})^{2}) 
\sim \mathcal{O}(G_{4}(m^{(m_{1}\cdots m_{p})}_{p,\,\text{winding}})^{2})$ by 
(\ref{e5}), (\ref{e15}) and (\ref{e47_2}). 
The density perturbation has been considered in various brane gas models 
\cite{bio_BW, bio_WB2, bio_W, bio_GP, bio_GP2, bio_NGP}.

In the brane world model, the wrapped branes distribute in a bulk space 
as well as on the brane expanding the three dimensions. 
Then, various bound states may arise between  many branes and 
the dark matter of wrapped branes may have interactions with the dark energy 
through the couplings of moduli fields. 
It may be interesting to investigate the dynamics of the universe with  the light branes in 
the four-dimensional Einstein frame for various models \cite{bio_GKP, bio_KKLT, 
bio_KKLMMT, bio_SL, bio_WT, bio_SS, bio_BEM, bio_BBEM,bio_SB, bio_BA2, bio_BDR, bio_FMR, bio_AK, bio_BM, bio_NO1, bio_NO2, bio_NO3, bio_NO4, bio_NO5}.

\begin{acknowledgments}
The authors thank the Yukawa Institute for Theoretical Physics at Kyoto University. 
Discussions during the YITP workshop YITP-W-08-04 on ``Development of Quantum Field Theory and String Theory'' 
were useful to complete this work. 
We also thank Particle Theory Group of the Yukawa Institute for Theoretical Physics for fruitful discussions. 
We would like to thank Tohru Eguchi, Kenji Hotta, Tetsuji Kimura, Hideo Kodama, Shinji Mukohyama, Misao Sasaki, 
Naoki Sasakura, Seiji Terashima  for useful and helpful comments.  
M.S. is supported by Sasagawa Scientific Research Grant from The Japan Science Society 
and by Hokkaido University Clark Memorial Foundation.
\end{acknowledgments}

\appendix

\section{$(d+1)$-dimensional Einstein frame} \label{appendixA}

We will decompose the ten-dimensional space-time as $10=(d+1)+(10-(d+1))$ and take the $(d+1)$-dimensional Einstein frame.  
The (d+1)-dimensional Einstein frame is obtained by a toroidal compactification of the circumference of $2\pi \sqrt{\alpha'}e^{\lambda_{m}}$ as 
\begin{align}
&\frac{1}{16 \pi G_{10}} \int d^{10}X \sqrt{-G}e^{-2\phi} R \\ \notag 
&=\frac{1}{16 \pi G_{d+1}} \int d^{d+1}x \sqrt{-g_{d+1}} R_{d+1}+\cdots \label{ap1} 
\end{align}
where the $(d+1)$-dimensional gravitational constant is given by 
$G_{d+1}=G_{10}/(2\pi \sqrt{\alpha'})^{10-(d+1)}$. 
To obtain this effective action we will consider the following variables: 
\begin{equation}
\begin{split}
&G_{\mu\nu}=e^{2\beta_{d+1}} g_{d+1,\, \mu\nu}, \\
&G_{mn}=e^{2 \lambda_{m}}\delta_{mn}, \\
& \beta_{d+1}=\frac{2}{d-1}\phi-\frac{1}{d-1} \sum_{m=d+1}^{9}\lambda_{m},
\end{split} \label{app2}
\end{equation}
where ( $\mu,\,\nu=0,\,1,\,\cdots ,\, d$ ) and ( $m,\,n=d+1,\,d+2,\,\cdots ,\,9$ ). 
Eq.(\ref{app2}) defines the $(d+1)$-dimensional Einstein frame. 
If we consider $d=3$ and $\beta_{3+1}\equiv \beta$, the four-dimensional Einstein frame 
defined by (\ref{e3}) is recovered.

\section{S-duality in the four-dimensional Einstein frame}   \label{appendixB}

The S-duality rule is defined by the ten-dimensional Einstein frame. 
Using (\ref{app2}) in the case of $d=9$, we obtain 
\begin{equation}
\begin{split}
G_{AB}&=e^{2\beta_{10}} g_{10,\, AB}, \\
\beta_{10}&=\frac{1}{4}\phi=\frac{1}{4} \Biggl( \beta+\frac{\overline{\lambda}}{2} \Biggr)
\end{split} \label{app3}
\end{equation}
where ($A,\,B=0,\,1,\,\cdots ,\,9$). $\beta_{10}$ 
is related with variables ($\beta,\, \overline{\lambda}$) which are defined by (\ref{e3}) as (\ref{app3}). 
Then the S-duality is given by $\phi \rightarrow -\phi$ which means that 
the string-frame metric transforms as 
\begin{equation}
G_{AB}\longrightarrow e^{-\phi} G_{AB}.  \label{app4}
\end{equation} 
to find the transformation rule of the S-duality in the four-dimensional Einstein frame 
we will assume 
\begin{equation}
\begin{split}
\lambda_{m}&\rightarrow \lambda_{m}+f_{\text{S-dual}}, \\
\beta  &\rightarrow \beta  +f_{\text{S-dual}}. 
\end{split}\label{app5}
\end{equation} 
If this transformation represents the S-duality satisfying (\ref{app4}), the action (\ref{e4}) 
must be invariant under this transformation. Substituting (\ref{app5}) for (\ref{e4}), 
the the S-dual invariance requires 
\begin{equation}
f_{\text{S-dual}}=-\frac{1}{2}\phi.   \label{app6}
\end{equation}
It is found that this solution satisfies (\ref{app4}), substituting  (\ref{app6}) for (\ref{app3}).

\section{fundamental string gas}  \label{appendixC}

 If the fundamental string wraps over a cycle of $T^{6}$ and 
we choose $\sigma^{0}=t$, $0 \leq \sigma^{1} \leq 2 \pi \sqrt{\alpha'}$,  
we can assume 
\begin{equation}
\begin{split}
X^{0}&=t,  \\
X^{m_{a}}&=X^{m_{a}}(t)+w^{m_{a}}\sigma^{1}, \quad (a=1,\,2,\,3,\,\cdots ,\,6),
\end{split}  \label{app7}
\end{equation} 
where we have considered the only zero mode and $w^{m_{a}}$ indicates the winding number of the string. 
The above assumption is motivated by $X^{m_{a}}(t,\,\sigma^{1}+2\pi \sqrt{\alpha'})=X^{m_{a}}(t,\,\sigma^{1})+2\pi \sqrt{\alpha'} w^{m_{a}}$.
We do not consider a dependence of $X^{i}$ ($i=1,\,2,\,3$), for simplicity.
The induced metric is given by 
\begin{equation}
\begin{split}
\gamma_{00}&= -e^{2\lambda_{0}}+\sum_{a=1}^{6}e^{2\lambda_{m_{a}}} (\dot{X}^{m_{a}})^{2},\\
\gamma_{01}&=  \sum_{a=1}^{6}e^{2\lambda_{m_{a}}}\dot{X}^{m_{a}} w^{m_{a}},\\
\gamma_{11}&=\sum_{a=1}^{6}e^{2\lambda_{m_{a}}}(w^{m_{a}})^{2}.
\end{split}  \label{app8}
\end{equation}
We will impose $\gamma_{0m_{1}}=0$, the meaning of which is explained later. 
Then the world volume action of a fundamental string wrapping over a cycle of $T^{6}$ is given by 
\begin{align}
S_{\text{F}1}&=-T_{\text{F}1} \int_{R \times \varSigma_{1}} d\sigma^{0} d\sigma^{1} \sqrt{-\gamma}  \notag \\
&=-(2 \pi \sqrt{\alpha'})T_{\text{F}1} \int dt e^{\lambda_{0}} \notag \\ 
& \times  \{\sum_{a=1}^{6}e^{2\lambda_{m_{a}}}(w^{m_{a}})^{2} (1
-\sum_{a=1}^{6}e^{2\lambda_{m_{a}}-2\lambda_{0}}(\dot{X}^{m_{a}})^{2}  )\}^{\frac{1}{2}}   \notag \\
&=-(2 \pi \sqrt{\alpha'})T_{\text{F}1} \int dt e^{\beta+n} \notag \\ 
& \times  \{\sum_{a=1}^{6}e^{2\lambda_{m_{a}}}(w^{m_{a}})^{2} (1
-\sum_{a=1}^{6}e^{2\lambda_{m_{a}}-2\beta-2n}(\dot{X}^{m_{a}})^{2} )  \}^{\frac{1}{2}}  \label{app9}
\end{align} 
where $T_{\text{F}1}=T_{\text{D}1}$. The equation of motion is 
\begin{align}
 &(\sum_{a=1}^{6}e^{2\lambda_{m_{a}}}(w^{m_{a}})^{2})^{1/2} e^{n+\beta} e^{2\lambda_{m_{a}}-2\beta-2n} \dot{X}^{m_{a}}\notag \\
 &=n_{m_{a}}
 \{1-\sum_{a=1}^{6}e^{2\lambda_{m_{a}}-2\beta-2n}(\dot{X}^{m_{a}})^{2}  \}^{\frac{1}{2}} \label{app10}
\end{align}
where $n_{m_{a}}$ is a constant of integration. 
By the definition of the energy-momentum tensor (\ref{e13}) and (\ref{app10}), we obtain 
\begin{align}
u_{\text{F}1}&=e^{-3A} m_{\text{F}1} \notag \\
&=e^{-3A}(2\pi \sqrt{\alpha'})T_{\text{F}1}  \notag  \\
& \times \Biggl\{ \sum_{a=1}^{6}e^{2\beta+2\lambda_{m_{a}}}(w^{m_{a}})^{2}
+ \sum_{a=1}^{6}e^{2\beta-2\lambda_{m_{a}}}(n_{m_{a}})^{2} \Biggr\}^{\frac{1}{2}}.   \label{app11} 
\end{align}
Substituting (\ref{app10}) for $\gamma_{0m_{1}}=0$, we obtain 
\begin{equation}
\sum_{a=1}^{6}n_{m_{a}}w^{m_{a}}=0.  \label{app12}
\end{equation}
If $n_{m_{a}}$ and $w^{m_{a}}$ are integers, (\ref{app12}) indicates the level matching condition, $L_{0}-\tilde{L}_{0}=0$, for the only zero mode.  
The mass $m_{\text{F}1}$ is invariant under the T-duality, ( $\lambda_{m_{a}}\rightarrow -\lambda_{m_{a}}$, $\phi\rightarrow \phi-\lambda_{m_{a}}$ )
with ( $n_{m_{a}} \longleftrightarrow  w^{m_{a}}$ ).

\section{K$\ddot{\text{A}}$hler form}  \label{appendixD}

The K$\ddot{\text{a}}$hler form $J$ is expanded by a harmonic form $\omega_{A}$ which is a basis of $H^{1,1}(\text{CY}_{3})$ as
\begin{equation}
J=v^{A}\omega_{A}. \label{appD1}
\end{equation}
where ($A=1,\,\cdots ,\,h^{1,1}$). 
It is useful to define the following quantities: 
\begin{equation}
\begin{split}
\mathcal{K}_{ABC}&=\int_{\text{CY}_{3}}\omega_{A} \wedge \omega_{B} \wedge \omega_{C}, \\
\mathcal{K}_{AB}&=\int_{\text{CY}_{3}}\omega_{A} \wedge \omega_{B} \wedge J=\mathcal{K}_{ABC}v^{C}, \\
\mathcal{K}_{A}&=\int_{\text{CY}_{3}}\omega_{A} \wedge J\wedge J=\mathcal{K}_{ABC}v^{B}v^{C}, \\  
\mathcal{K}&=\int_{\text{CY}_{3}}J\wedge J\wedge J=\mathcal{K}_{ABC}v^{A}v^{B}v^{C}=6V_{\text{CY}_{3}}.
\end{split}  \label{appD2}
\end{equation}
Using the following relation \cite{bio_ST2}:   
\begin{equation}
\begin{split}
*\omega_{A}&=-J\wedge \omega_{A}+\frac{3}{2}\frac{\mathcal{K}_{A}}{\mathcal{K}}J\wedge J,  \\
*J&=\frac{1}{2} J\wedge J, 
\end{split}  \label{appD3}
\end{equation}
we obtain 
\begin{equation}
\int_{\text{CY}_{3}}\omega_{A}\wedge *\omega_{B}=-\mathcal{K}_{AB}+\frac{3}{2}\frac{\mathcal{K}_{A}\mathcal{K}_{B}}{\mathcal{K}}. \label{appD4}
\end{equation}
Eq.(\ref{appD4}) is related to gauge couplings of the vector multiplets in the low energy effective action of 
type IIA supergravity compactified on a Calabi-Yau threefold. 
If the NSNS 2-form $B_{2}$ vanishes, the matrix of the gauge couplings is given by \cite{bio_GL, bio_GL2, bio_LM, bio_BGHL}
\begin{equation}
\begin{split}
\text{Im}\mathcal{N}_{00}&=-\frac{\mathcal{K}}{6}=-V_{\text{CY}_{3}}, \\
\text{Im}\mathcal{N}_{AB}&=-\int_{\text{CY}_{3}}\omega_{A}\wedge *\omega_{B}. 
\end{split} \label{appD5}
\end{equation}
Introducing $\mathcal{K}^{AB}$ by $\mathcal{K}^{AB}\mathcal{K}_{AC}=\delta^{A}_{~C}$, one derives the inverse matrix: 
\begin{equation}
\begin{split}
(\text{Im}\mathcal{N}^{-1})^{00}&=-\frac{6}{\mathcal{K}},  \\
(\text{Im}\mathcal{N}^{-1})^{AB}&=-\int_{\text{CY}_{3}}\tilde{\omega}^{A}\wedge *\tilde{\omega}^{B} 
\end{split} \label{appD6}
\end{equation}
where we defined the dual basis $\tilde{\omega}^{A}\in H^{2,2}(\text{CY}_{3})$ by
\begin{equation}
\int_{\text{CY}_{3}}\omega_{A}\wedge \tilde{\omega}^{B}=\delta^{B}_{~A}  \label{appD7}
\end{equation}
and 
\begin{equation}
\begin{split}
\int_{\text{CY}_{3}}\tilde{\omega}^{A}\wedge * \tilde{\omega}^{B}&=-\mathcal{K}^{AB}+\frac{3v^{A}v^{B}}{\mathcal{K}}, \\
* \tilde{\omega}^{A}&=\Biggl( -\mathcal{K}^{AB}+\frac{3v^{A}v^{B}}{\mathcal{K}} \Biggr) \omega_{B}. 
\end{split}
\end{equation}

\section{Hodge dual of ($\alpha_{\hat{K}},\,\beta^{\hat{K}}$) }  \label{appendixE}

$\{ \alpha_{\hat{K}},\,\beta^{\hat{K}} \}$  are both three-forms and those Hodge duals are also three-forms which 
 can be expanded in term of $\alpha_{\hat{K}}$ and $\beta^{\hat{K}}$ according to 
\begin{align}
\begin{split}
&*\alpha_{\hat{K}}=A_{\hat{K}}^{~~\hat{L}}\alpha_{\hat{L}}+B_{\hat{K}\hat{L}}\beta^{\hat{L}}, \\
&*\beta^{\hat{K}}=C^{\hat{K}\hat{L}}\alpha_{\hat{L}}+D^{\hat{K}}_{~~\hat{L}}\beta^{\hat{L}}.
\end{split}  \label{app13}
\end{align}
Using 
\begin{equation}
\int_{\text{CY}_{3}} \alpha_{\hat{K}}\wedge \beta^{\hat{L}}=\delta^{\hat{L}}_{\hat{K}},  \notag 
\end{equation}
one derive 
\begin{align}
B_{\hat{K}\hat{L}}&=\int_{\text{CY}_{3}} \alpha_{\hat{K}} \wedge * \alpha_{\hat{L}}
=\int_{\text{CY}_{3}} \alpha_{\hat{L}} \wedge * \alpha_{\hat{K}}=B_{\hat{L}\hat{K}},  \notag \\
-C^{\hat{K}\hat{L}}&=\int_{\text{CY}_{3}} \beta^{\hat{K}} \wedge * \beta^{\hat{L}}
=\int_{\text{CY}_{3}} \beta^{\hat{L}} \wedge * \beta^{\hat{K}}=-C^{\hat{L}\hat{K}}, \notag  \\
-A_{\hat{K}}^{~~\hat{L}}&=\int_{\text{CY}_{3}} \beta^{\hat{L}} \wedge * \alpha_{\hat{K}}
=\int_{\text{CY}_{3}} \alpha_{\hat{K}} \wedge * \beta^{\hat{L}}=D^{\hat{L}}_{~~\hat{K}}.  \notag  
\end{align}
The matrices $A$, $B$, $C$, $D$ are determined in term of a matrix $\mathcal{M}$ \cite{bio_H, bio_CDF}
\begin{equation}
\begin{split}
&A=(\text{Re}\mathcal{M})(\text{Im}\mathcal{M})^{-1},  \\
&B=-(\text{Im}\mathcal{M})-(\text{Re}\mathcal{M})(\text{Im}\mathcal{M})^{-1}(\text{Re}\mathcal{M}), \\
&C=(\text{Im}\mathcal{M})^{-1}, 
\end{split}  \label{app15}
\end{equation}
where 
\begin{equation}
\mathcal{M}_{\hat{K}\hat{L}}=\overline{\mathcal{F}}_{\hat{K}\hat{L}}+2i 
\frac{(\text{Im}\mathcal{F})_{\hat{K}\hat{M}}Z^{\hat{M}} (\text{Im}\mathcal{F})_{\hat{L}\hat{N}}Z^{\hat{N}}}
{Z^{\hat{M}}(\text{Im}\mathcal{F})_{\hat{M}\hat{N}}Z^{\hat{N}}}.  \label{app16}
\end{equation}
The matrix $\mathcal{M}$ satisfies 
\begin{equation}
\mathcal{F}_{\hat{K}}=\mathcal{M}_{\hat{K}\hat{L}}Z^{\hat{L}}  \label{app17}
\end{equation}
where we have used the following relations on a holomorphic 
prepotential $\mathcal{F}$ with respect to $Z^{\hat{K}}$: 
\begin{equation}
\begin{split}
\mathcal{F}&=\frac{1}{2}Z^{\hat{K}}\mathcal{F}_{\hat{K}}, \\
\mathcal{F}_{\hat{K}}&=\frac{\partial \mathcal{F}}{\partial Z^{\hat{K}}}=Z^{\hat{K}}\mathcal{F}_{\hat{K}\hat{L}}, \\
\mathcal{F}_{\hat{K}\hat{L}}&=\frac{\partial \mathcal{F}_{\hat{L}}}{\partial Z^{\hat{K}}}. \\ 
\end{split}  \label{app18}
\end{equation}
\\
\\

\end{document}